\documentclass{llncs}
\usepackage{etex}
\usepackage{float}
\usepackage{wrapfig}
\usepackage{longtable}
\usepackage{supertabular}

\usepackage{amsmath}
\usepackage{amssymb}
\usepackage{stmaryrd}

\usepackage[all]{xy}
\usepackage {mathpartir}
\usepackage{graphicx}
\usepackage{url}
\usepackage{tikzfig}
\tikzstyle{every picture}=[baseline=-0.25em]
\tikzstyle{dotpic}=[scale=0.6]
\tikzstyle{diredges}=[every to/.style={diredge}]
\tikzstyle{dot graph}=[shorten <=-0.1mm,shorten >=-0.1mm,scale=0.6]
\tikzstyle{digraph}=[-latex]
\tikzstyle{plot point}=[circle,fill=black,minimum width=2mm,inner sep=0]
\tikzstyle{string graph}=[scale=0.6]
\tikzstyle{sg diredge}=[-stealth]
\tikzstyle{rewrite edge}=[-open triangle 45]
\tikzstyle{sg bold diredge}=[-stealth,thick,shorten >=-1pt]
\tikzstyle{sg vertex}=[circle,minimum width=2.2mm,fill=white,draw=black,inner sep=0mm]
\tikzstyle{labelled sg vertex}=[circle,minimum width=7mm,fill=white,draw=black,inner sep=0mm]
\tikzstyle{small sg vertex}=[circle,minimum width=4mm,fill=white,draw=black,inner sep=0mm, font=\footnotesize]
\tikzstyle{sg grey vertex}=[sg vertex,fill=gray!30!white]
\tikzstyle{sg white vertex}=[sg vertex,fill=white]
\tikzstyle{sg black vertex}=[sg vertex,fill=black]
\tikzstyle{sg bold vertex}=[circle,minimum width=2.2mm,fill=white,draw=black,very thick,inner sep=0mm]
\tikzstyle{sg wire vertex}=[circle,minimum width=1mm,fill=black,inner sep=0mm]
\tikzstyle{cloud vertex}=[fill=white, draw=black, inner sep=2 mm, shape=cloud, aspect=1.5]
\tikzstyle{tick vertex}=[rectangle,fill=black,minimum height=1mm,minimum width=2.5mm,inner sep=0mm]

\tikzstyle{braceedge}=[decorate,decoration={brace,amplitude=2mm,raise=-1mm}]
\tikzstyle{small braceedge}=[decorate,decoration={brace,amplitude=1mm,raise=-1mm}]
\tikzstyle{left hook arrow}=[left hook-latex]
\tikzstyle{right hook arrow}=[right hook-latex]

\tikzstyle{bbox edge}=[draw=blue]
\tikzstyle{bbox include}=[->,draw=blue]
\tikzstyle{bbox corner}=[inner sep=0pt,rectangle,fill=blue,draw=blue,minimum width=1.5mm,minimum height=1.5mm]

\tikzstyle{west wire label}=[font=\footnotesize\it,anchor=west,inner sep=1pt,xshift=-3pt]
\tikzstyle{east wire label}=[font=\footnotesize\it,anchor=east,inner sep=1pt,xshift=3pt]

\tikzstyle{dot}=[inner sep=0.7mm,minimum width=0pt,minimum height=0pt,fill=black,draw=black,shape=circle]
\tikzstyle{white dot}=[dot,fill=white]
\tikzstyle{alt white dot}=[white dot,label={[xshift=2.9mm,yshift=-0.1mm]left:$\cdot$}]
\tikzstyle{gray dot}=[dot,fill=gray!50]
\tikzstyle{box vertex}=[draw=black,rectangle]
\tikzstyle{whitebg}=[fill=white,inner sep=2pt]
\tikzstyle{graph state vertex}=[sg vertex,fill=black]

\tikzstyle{square box}=[rectangle,fill=white,draw=black,minimum height=6mm,minimum width=6mm]
\tikzstyle{small square box}=[rectangle,fill=white,draw=black,minimum height=0mm,minimum width=8mm,inner sep=3pt]
\tikzstyle{square gray box}=[rectangle,fill=gray!30,draw=black,minimum height=6mm,minimum width=6mm]
\tikzstyle{point}=[regular polygon,regular polygon sides=3,draw=black,scale=0.75,inner sep=-0.5pt,minimum width=7mm,fill=white]
\tikzstyle{copoint}=[point,regular polygon rotate=180,fill=white]
\tikzstyle{gray point}=[point,fill=gray!40!white]
\tikzstyle{gray copoint}=[copoint,fill=gray!40!white]

\tikzstyle{open graph}=[baseline=-0.25em]
\tikzstyle{greybg}=[background rectangle/.style={fill=black!5,draw=black!30,rounded corners=1ex}, show background rectangle]

\tikzstyle{edge point}=[circle,minimum width=1mm,fill=black,inner sep=0mm]
\tikzstyle{vertex point}=[circle,minimum width=2.2mm,fill=white,draw=black,inner sep=0mm]
\tikzstyle{gray vertex point}=[circle,minimum width=2.2mm,fill=gray!30!white,draw=black,inner sep=0mm]
\tikzstyle{edge label}=[inner sep=2pt, font=\small]
\tikzstyle{on edge label}=[fill=white, font=\footnotesize, inner sep=1 pt]

\newcommand{\edgearrow}{{\arrow[black]{>}}}
\newcommand{\edgetick}{{\arrow[black,scale=0.7,very thick]{|}}}
\tikzstyle{diredge}=[->]
\tikzstyle{gray edge}=[gray!50!white]
\tikzstyle{medium diredge}=[->]
\tikzstyle{short diredge}=[->]
\tikzstyle{halfedge}=[-)]
\tikzstyle{other halfedge}=[(-]
\tikzstyle{freeedge}=[(-)]
\tikzstyle{white edge}=[line width=5pt,white]
\tikzstyle{tick}=[postaction=decorate,decoration={markings, mark=at position 0.5 with \edgetick}]
\tikzstyle{small map edge}=[|-latex, gray!60!blue, shorten <=0.9mm, shorten >=0.5mm]
\tikzstyle{thick dashed edge}=[very thick,dashed,gray!40]
\tikzstyle{map edge}=[|-latex,very thick, gray!40, shorten <=1mm, shorten >=0.5mm]
\tikzstyle{tickedge}=[postaction=decorate,
  decoration={markings, mark=at position 0.5 with \edgetick}]
\tikzstyle{dirtickedge}=[postaction=decorate,
  decoration={markings, mark=at position 0.5 with \edgetick},
  decoration={markings, mark=at position 0.85 with \edgearrow}]
\tikzstyle{dirdoubletickedge}=[postaction=decorate,
  decoration={markings, mark=at position 0.4 with \edgetick},
  decoration={markings, mark=at position 0.6 with \edgetick},
  decoration={markings, mark=at position 0.85 with \edgearrow}]

\tikzstyle{arrs}=[-latex,font=\small,auto]
\tikzstyle{arrow plain}=[arrs]
\tikzstyle{arrow dashed}=[dashed,arrs]
\tikzstyle{arrow bold}=[very thick,arrs]
\tikzstyle{arrow hide}=[draw=white!0,-]
\tikzstyle{arrow reverse}=[latex-]
\tikzstyle{cdnode}=[]

\tikzstyle{cnot}=[fill=white,shape=circle,inner sep=-1.4pt]
\tikzstyle{bang box}=[draw=black,dashed,minimum height=12mm,minimum width=12mm,fill=gray!20]
\tikzstyle{wire label}=[font=\footnotesize, auto]

\usepackage{xspace}
\graphicspath{{figures/}}

\usepackage{algpseudocode}
\usepackage{algorithm}

\newcommand{\cmdrewritesto}{\tikz[baseline=-0.25em] { \draw [-open triangle 45, line width=0.2pt] (0,0) -- (0.5,0); }\,}

\DeclareMathOperator{\rewritesto}{\cmdrewritesto}

\tikzstyle{cdiag}=[matrix of math nodes, row sep=2em, column sep=2em, text height=1.5ex, text depth=0.25ex,inner sep=0.5em]
\tikzstyle{arrow above}=[transform canvas={yshift=0.5ex}]
\tikzstyle{arrow below}=[transform canvas={yshift=-0.5ex}]

\definecolor{light-gray}{gray}{0.75}
\newcommand{\ncbox}[1]{\fcolorbox{light-gray}{light-gray}{#1}} 	
\newcommand{\desc}[1]{\small{~~\ncbox{#1}}}

\begin{document}
 
\newcommand{\grewr}[2]{\mbox{$#1 \multimap{} #2$}}
\newcommand{\tensor}{\otimes}
\newcommand{\grewrside}[3]{\mbox{$#1 \stackrel{#3}{\multimap{}}{} #2$}}

\newcommand{\todo}[1]{{\begin{center}\fbox{\begin{minipage}{0.95\linewidth}\textbf{TODO:} #1\end{minipage}}\end{center}}}
 \newcommand{\subwire}{\mbox{$\,<:\,$}}
  \newcommand{\notsubwire}{\mbox{$\not$\hspace{-3pt}}\subwire{}}
 \newtheorem{declaration}{Declaration}
 \newtheorem{assumption}{Assumption}

\newcommand{\grel}{\mbox{$\Downarrow$}}
\newcommand{\arrel}{\mbox{$\Downarrow$}}
\newcommand{\trel}{\mbox{$\Downarrow$}}
\newcommand{\outrel}{\mbox{$\Downarrow$}}

\newcommand{\evaltac}[1]{\mbox{$\Downarrow_{#1}^{tac}$}}
\newcommand{\evalstep}[1]{\mbox{$\Downarrow_{#1}^{step}$}}
\newcommand{\evalgraph}[1]{\mbox{$\Downarrow_{#1}^{*}$}}

\newcommand{\evalsimplestep}{\mbox{$~\Downarrow_{G}^{1}~$}}
\newcommand{\evalsimpleclosure}{\mbox{$~\Downarrow_{G}^{*}~$}}

\newcommand{\evalrecstep}{\mbox{$~\Downarrow_{rs}^{R1}~$}}
\newcommand{\evalrecbranchstep}{\mbox{$~\Downarrow_{rs}^{RBr}~$}}
\newcommand{\evalrecclosure}{\mbox{$~\Downarrow_{rs}^{R*}~$}}

\newcommand{\leftsq}{\textrm{[}}
\newcommand{\rightsq}{\textrm{]}}

\renewcommand{\todo}[1]{}

\newcommand{\beforesection}{\vspace{-0.6em}}
\newcommand{\aftersection}{\vspace{-0.4em}}

\mainmatter              
\title{A Graphical Language for Proof Strategies}
\titlerunning{A Graphical Language for Proof Strategies}  
%
\author{Gudmund Grov\inst{1}, Aleks Kissinger\inst{2} and Yuhui Lin\inst{1}}
\authorrunning{G. Grov, A. Kissinger and Y. Lin} 
%
\tocauthor{Gudmund Grov, Aleks Kissinger and Yuhui Lin}
\institute{School of Mathematical and Computer Sciences, 
Heriot-Watt University, Edinburgh, UK,
\email{\{G.Grov,Y.Lin\}@hw.ac.uk}
\and
Department of Computer Science, University of Oxford, UK,
\email{aleks.kissinger@cs.ox.ac.uk}}

\maketitle              

\begin{abstract}
Complex automated proof strategies are often difficult to extract, visualise, modify, and debug. 
Traditional tactic languages, often based on stack-based goal propagation, make it easy to write proofs that obscure the flow of goals between tactics and are fragile to minor changes in input, proof structure or changes to tactics themselves. Here, we address this by introducing a graphical language called PSGraph for writing proof strategies. Strategies are constructed visually by ``wiring together'' collections of tactics and evaluated by propagating goal nodes through the diagram via graph rewriting. Tactic nodes can have many output wires, and use a filtering procedure based on goal-types (predicates describing the features of a goal) to decide where best to send newly-generated sub-goals. 
In addition to making the flow of goal information explicit, the graphical language can fulfil the role of many tacticals using visual idioms like branching, merging, and feedback loops. We argue that this language enables development of more robust proof strategies and provide several examples, along with a prototype implementation in Isabelle.
\end{abstract}

\vspace{-0.6em}
\beforesection
\section{Introduction}\label{sec:intro}
\aftersection

Most tactic languages for interactive theorem provers are not designed to distinguish goals in cases where tactics
produce multiple sub-goals. Thus when composing tactics, one  has no choice but to rely on the order in which goals
arrive, thus making them brittle to minor changes.  For example, consider a case where we expect three sub-goals from
tactic $t_1$, where the first two are sent to $t_2$ and the last to $t_3$. A small improvement of $t_1$ may result in
only two sub-goals. This ``improvement'' causes $t_2$ to be applied to the second goal when it should have been $t_3$.
The tactic $t_2$ may then fail or create unexpected new sub-goals that cause some later tactic to fail.

As a result: (1) it is often difficult to compose tactics in such a way that all sub-goals are sent to the correct
target tactic, especially when different goals should be handled differently; (2) when a large tactic fails, it is hard
to analyse where the failure occurred; and (3) the reliance of goal order means that machine learning new tactics from existing
proofs have not been as successful for tactics as it has been for discovering relevant hypothesis in automated theorem
provers.

Moreover, if the structure of a tactic is difficult to understand, often the easiest way for a user to
deal with failure is to manually guide the proof until the tactic succeeds (or becomes unnecessary), rather
than correcting the weakness of the tactic itself. In this case, the proof is made more complicated and
insight from this failure is not carried across to other proofs. Thus, a tactic language where it is easy
to diagnose and correct failures will lead to better tactics and simpler, more general proofs.

This can be achieved in part by attempting to find as many errors as possible \textit{statically}. The problem with
existing tactic languages is that tactics are essentially untyped: they are essentially functions from a \emph{goal} to
a conjunction of \emph{sub-goals}. In many programming languages, types are used statically to rule out many ``obvious''
errors. For example, in typed functional languages, a type error will occur when one tries to compose two functions
which do not have a unifiable type. In an untyped tactic language, this kind of ``round-peg-square-hole'' situation will
not manifest until run-time.

For errors that cannot be found statically, it is very hard to inspect and analyse the failures during debugging. In the
above example, if $t_2$ creates sub-goals that tactics  later in the proof do not expect, the error may be reported in a
completely different place. Without a clear handle on the flow of goals through the proof, finding the real source of
the error could be very difficult indeed.

In this paper, we address these issues by introducing a graphical proof strategy language called \emph{PSGraph}. We
argue that this language has three advantages over more traditional tactic languages: (i) it improves robustness of
proof strategies with static goal typing and type-safe tactic ``wirings''; (ii) it improves the ability to dynamically
inspect, analyse, and modify strategies, especially when things go wrong; and (iii) it enables machine learning of new tactics
from proofs.

For the sake of this paper, we shall focus on (i) and (ii). A discussion on the use of PSGraph for (iii) can be found
in~\cite{grov13a}, where a form of of \textit{analogous reasoning} through tactic generalisation is developed using
PSGraph.

A high-level introduction to PSGraph is given Section \ref{sec:psgraph}, followed by a discussion on goal types in Section \ref{sec:tactic}. Section \ref{sec:lang} gives a detailed description of the language and evaluation, before combinators and hierarchies are introduced in Section \ref{sec:graph-tactics}. An Isabelle implementation, including experiments, is given in Section \ref{sec:impl}. We then discuss related work (Section \ref{sec:related}) and conclude (Section \ref{sec:conc}).

\beforesection
\section{Proof Strategy Graphs = Tactics + Plumbing}\label{sec:psgraph}
\aftersection

A useful analogy for thinking about designing sophisticated tactics is that of plumbing. Instead of thinking of tactics as functions that compose, think of them as individual components whose inputs and outputs can be connected by various pipes. Each component of the system is a tactic of the underlying theorem prover, and your job in designing a proof strategy is to create a network of tactics by plugging input and output from tactics together.

In a pipe network, pipes comes in all sizes and shapes, and you can only connect the same \emph{type} of pipes together -- after all, there is a reason you don't connect the toilet waste water to the mains water. The same is true for tactics: they only work for certain goals (although for some tactics this range of goals is rather wide). For example, an `assumption' tactic expects a hypothesis to be unifiable with the goal, and `$\forall$-intro' expects the goal to start with a $\forall$ quantifier. 

\begin{figure}
  \centering
  \vspace{-12pt}
  \scalebox{0.8}{%
\beginpgfgraphicnamed{string_diagram}
\begin{tikzpicture}[string graph]
	\begin{pgfonlayer}{nodelayer}
		\node [style=none] (0) at (-0.75, 3) {};
		\node [style=none] (1) at (0.25, 3) {};
		\node [style=none] (2) at (-0.75, 2) {};
		\node [style=none] (3) at (0.25, 2) {};
		\node [style=square box, minimum width=1.5 cm] (4) at (-0.25, 1.5) {$f$};
		\node [style=none] (5) at (-0.75, 1.25) {};
		\node [style=none] (6) at (0.25, 1.25) {};
		\node [style=none] (7) at (-1.75, 0) {};
		\node [style=square box, minimum width=1.5 cm] (8) at (-1.75, -0.5) {$g$};
		\node [style=none] (9) at (-2.25, -0.75) {};
		\node [style=none] (10) at (-1.25, -0.75) {};
		\node [style=none] (11) at (-2, -2) {};
		\node [style=none] (12) at (-0.5, -2) {};
		\node [style=square box, minimum width=1.5 cm] (13) at (-1.25, -2.5) {$h$};
		\node [style=none] (14) at (0.75, -2.25) {};
		\node [style=none] (15) at (0.75, -3) {};
		\node [style=none, wire label] (16) at (-1.25, 2.5) {$A$};
		\node [style=none, wire label] (17) at (0.75, 2.5) {$B$};
		\node [style=none, wire label] (18) at (-2, 0.5) {$C$};
		\node [style=none, wire label] (19) at (0.5, 0) {$D$};
		\node [style=none, wire label] (20) at (1.25, -2.5) {$E$};
		\node [style=none, wire label] (21) at (-2.75, -1.5) {$A$};
		\node [style=none] (22) at (2.25, 3) {};
		\node [style=none] (23) at (2.25, -3) {};
		\node [style=none, wire label] (24) at (2.75, 1) {$C$};
	\end{pgfonlayer}
	\begin{pgfonlayer}{edgelayer}
		\draw [style=diredge, in=90, out=-90] (6.center) to (12.center);
		\draw [in=90, out=-90] (10.center) to (14.center);
		\draw [style=diredge, in=90, out=-90] (5.center) to (7.center);
		\draw [style=diredge] (0.center) to (2.center);
		\draw [style=diredge] (14.center) to (15.center);
		\draw [style=diredge, in=90, out=-90] (9.center) to (11.center);
		\draw [style=diredge] (1.center) to (3.center);
		\draw [style=diredge, in=90, out=-90] (22.center) to (23.center);
	\end{pgfonlayer}
\end{tikzpicture}}
\endpgfgraphicnamed}
  \caption{A string diagram}\label{fig:string-diagram}
  \vspace{-14pt}
\end{figure}
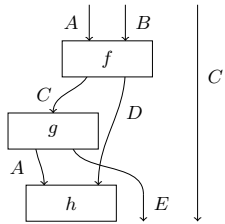

Formally, we represent a ``pipe network'' as a \textit{string diagram} (see Fig. \ref{fig:string-diagram}) \cite{paper:Dixon:10}, and we represent dynamics, or ``goals flowing down pipes'' using string diagram rewriting. String diagrams consist of \textit{boxes}, representing processes and typed \textit{wires} that connect them together. Unlike graph edges, wires need not be connected to a box at both ends, but can be left open to represent inputs and outputs. Just like a piece of pipe on its own, a wire that is open at both ends represents the \textit{identity} or ``do-nothing'' process.

A \textit{string diagram rewrite rule} is a pair of string diagrams $L$ and $R$ sharing the same boundary (i.e. there are type-respecting bijections of the respective inputs and outputs). Typically we write this $L \rewritesto R$. In order to apply a rewrite rule, one first finds a \textit{matching} $m : L \to G$, which is an embedding of $L$ into $G$ respecting the type of wires and the input/output arities of boxes. Once a matching is found, the image of $m$ is cut out of $G$ and replaced with $R$ to produce a new graph. The fact that there exists a bijection of the boundary between $L$ and $R$ is crucial to the final step, because it tells us precisely how to ``glue'' $R$ into the location that $L$ used to be. This agrees with a visual intuition for diagram substitution, and can also be formalised using double-pushout graph rewriting. For details, see~\cite{paper:Dixon:10}.

Proof strategy graphs (PSGraphs) are string diagrams whose boxes are labelled with tactics. As with the plumbing analogy, we think the typing information associated with a pipe as a property of the pipe itself. For that reason, we label wires with \textit{goal types}, which are predicates defined on goals. Intuitively, these provide information about some characteristics, such as ``shape'', of a goal, which are used to influence the path a goal takes as it passes through the strategy graph. To represent a goal being on a wire, we introduce a special \emph{goal} node to the graph. In the diagrams, we draw such nodes as a circle, while a tactic is a rectangle. 

One evaluation step works by a single tactic node on a single goal. Here, the goal is consumed  from the input wire, the tactic in the tactic node is applied to the goal, and the resulting sub-goals (if any) are sent down the output wires where they match. When all the goal nodes are in the output wires of the graph, i.e. a wire with an open destination, then it has successfully evaluated.
If no output type matches a goal, then evaluation fails. For evaluation this improves robustness of the tactic in two ways: (1) since composition is over the \emph{type of goals}, we avoid the brittleness arising from defining composition in terms of the number of sub-goals or order of sub-goals, and (2) if an unexpected sub-goal arises then evaluation will fail at the actual point of failure as it will not fit into any of the output pipes. In general, we allow this evaluation procedure to be non-deterministic by introducing branching whenever a tactic behaves non-deterministically, or a sub-goal produced by a tactic matches more than one output wire. However, with appropriate choice of goal types and evaluation strategy, this branching can be minimised.


\begin{wrapfigure}[14]{r}{0.35\textwidth}
 \centering
 \vspace{-11mm}
 \scalebox{0.99}{%
\beginpgfgraphicnamed{induct_ex}
\begin{tikzpicture}[string graph]
	\begin{pgfonlayer}{nodelayer}
		\node [style=west wire label, fill=white, xshift=3pt, yshift=-1pt] (0) at (1.75, 2) {can-ripple};
		\node [style=none] (1) at (1.5, 1.25) {};
		\node [style=none] (2) at (-1.25, 1.25) {};
		\node [style=none] (3) at (1.5, 1.75) {};
		\node [style=square box] (4) at (-0.75, 3) {induct};
		\node [style=none] (5) at (1.5, 0.25) {};
		\node [style=none] (6) at (-1.25, -2.75) {};
		\node [style=none] (7) at (0.5, 1.25) {};
		\node [style=square box] (8) at (1, 0.75) {ripple};
		\node [style=none] (9) at (-0.25, 2.5) {};
		\node [style=none] (10) at (2.5, 0.25) {};
		\node [style=none] (11) at (0.5, -1.25) {};
		\node [style=none] (12) at (2.5, 1.75) {};
		\node [style=none] (13) at (-1.25, 0.5) {};
		\node [style=none] (14) at (0.5, 0.25) {};
		\node [style=none] (15) at (-1.25, 2.5) {};
		\node [style=none] (16) at (1.5, 0.25) {};
		\node [style=none] (17) at (-0.25, 2) {};
		\node [style=west wire label] (18) at (0, 2) {step};
		\node [style=west wire label] (19) at (0.75, -0.75) {rippled};
		\node [style=square box] (20) at (0.5, -1.75) {fertilise};
		\node [style=west wire label] (21) at (-0.5, -4.25) {any};
		\node [style=square box] (22) at (-0.75, -3.25) {simp};
		\node [style=none] (23) at (-0.75, -4.5) {};
		\node [style=none] (24) at (-0.75, -3.75) {};
		\node [style=none] (25) at (0.5, -2.25) {};
		\node [style=none] (26) at (-0.25, -2.75) {};
		\node [style=west wire label] (27) at (0.5, -2.75) {any};
		\node [style=west wire label] (28) at (-1, -0.25) {base};
		\node [style=west wire label] (29) at (-0.5, 4) {inductable};
		\node [style=none] (30) at (-0.75, 3.5) {};
		\node [style=none] (31) at (-0.75, 4.5) {};
	\end{pgfonlayer}
	\begin{pgfonlayer}{edgelayer}
		\draw [style=diredge, in=90, out=-90, looseness=1.25] (17.center) to (7.center);
		\draw [style=diredge, in=90, out=-90] (14.center) to (11.center);
		\draw [in=90, out=-90, looseness=1.25] (2.center) to (13.center);
		\draw [style=diredge] (13.center) to (6.center);
		\draw [in=-90, out=-90, looseness=2.00] (5.center) to (10.center);
		\draw (10.center) to (12.center);
		\draw [in=90, out=90, looseness=2.00] (12.center) to (3.center);
		\draw (16.center) to (5.center);
		\draw (9.center) to (17.center);
		\draw (15.center) to (2.center);
		\draw [style=diredge] (3.center) to (1.center);
		\draw [style=diredge] (24.center) to (23.center);
		\draw [style=diredge, in=79, out=-105, looseness=1.50] (25.center) to (26.center);
		\draw [style=diredge] (31.center) to (30.center);
	\end{pgfonlayer}
\end{tikzpicture}}
\endpgfgraphicnamed}
 \vspace{-2mm}
 \caption{Rippling}\label{fig:rippling}
\end{wrapfigure}
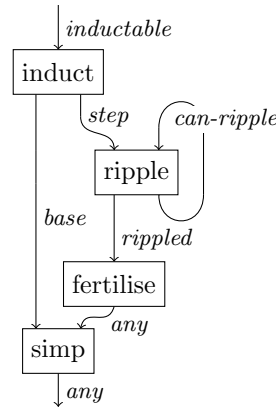

An example of a proof strategy which relies on specific properties of a goal is \emph{rippling}~\cite{rippling-book}. It is a rewriting technique most commonly used on step cases of inductive proofs. It ensures that each `ripple' step moves the goal towards the induction hypothesis (IH). This step is repeated until the IH can be applied to simplify or fully discharge the goal -- a process called `fertilisation'. The advantage of rippling is that it is guaranteed to terminate, whilst allowing rewriting behaviour that would not otherwise terminate (e.g. allowing a rewrite rule to be applied in both directions). Termination is ensured by checking that a certain \emph{embedding} property holds for the goal being rippled, while a measure is reduced from a previous goal. Collectively, these properties are captured by a goal type, in this cased called `\textit{can-ripple}'. When a goal is fully `rippled', then  `fertilisation' is applied. Fig. \ref{fig:rippling} illustrates a variant of ``induction with rippling" in PSGraph, where the base case and any resulting goals from the rippling process is sent to the `simp' tactic.

\begin{example}\label{ex:even1}
  Evaluating the top half of the strategy graph given in Fig. \ref{fig:rippling}:
  \begin{center}
    \scalebox{0.8}{%
\beginpgfgraphicnamed{rewrite_ex}
\begin{tikzpicture}[string graph]
	\begin{pgfonlayer}{nodelayer}
		\node [style=square box] (0) at (-9.25, -0.25) {ripple};
		\node [style=none] (1) at (-8.75, -1.25) {};
		\node [style=small sg vertex] (2) at (-10.5, 3) {a};
		\node [style=none] (3) at (-10, 1) {};
		\node [style=none] (4) at (-10.5, 3.5) {};
		\node [style=none] (5) at (-9.75, 0.25) {};
		\node [style=square box] (6) at (-10.5, 2) {induct};
		\node [style=none] (7) at (-9.75, -2.75) {};
		\node [style=none] (8) at (-11, -0.5) {};
		\node [style=none] (9) at (-11, -2.75) {};
		\node [style=none] (10) at (-10.5, 2.5) {};
		\node [style=none] (11) at (-11, 0.25) {};
		\node [style=none] (12) at (-9.75, -0.75) {};
		\node [style=none] (13) at (-7.75, 0.75) {};
		\node [style=none] (14) at (-8.75, 0.75) {};
		\node [style=none] (15) at (-7.75, -1.25) {};
		\node [style=none] (16) at (-7, 0) {$\rewritesto$};
		\node [style=none] (17) at (-8.75, -0.75) {};
		\node [style=none] (18) at (-10, 1.5) {};
		\node [style=none] (19) at (-11, 1.5) {};
		\node [style=none] (20) at (-8.75, 0.25) {};
		\node [style=none] (21) at (-5.75, -2.75) {};
		\node [style=none] (22) at (-4.5, -2.75) {};
		\node [style=none] (23) at (-5.25, 3.5) {};
		\node [style=none] (24) at (-3.5, 0.75) {};
		\node [style=none] (25) at (-5.75, 0.25) {};
		\node [style=none] (26) at (-3.5, -0.75) {};
		\node [style=small sg vertex] (27) at (-5.75, 0.25) {b};
		\node [style=none] (28) at (-4.5, 0.25) {};
		\node [style=none] (29) at (-3.5, 0.25) {};
		\node [style=none] (30) at (-1.75, 0) {$\rewritesto$};
		\node [style=none] (31) at (-2.5, -1.25) {};
		\node [style=square box] (32) at (-4, -0.25) {ripple};
		\node [style=square box] (33) at (-5.25, 2) {induct};
		\node [style=none] (34) at (-3.5, -1.25) {};
		\node [style=none] (35) at (-4.75, 0.75) {};
		\node [style=none] (36) at (-5.25, 2.5) {};
		\node [style=none] (37) at (-4.75, 1.5) {};
		\node [style=small sg vertex] (38) at (-4.75, 1) {d};
		\node [style=none] (39) at (-5.75, 1.5) {};
		\node [style=none] (40) at (-4.5, -0.75) {};
		\node [style=none] (41) at (-5.75, -0.5) {};
		\node [style=none] (42) at (-2.5, 0.75) {};
		\node [style=small sg vertex] (43) at (-0.5, -1.25) {c};
		\node [style=none] (44) at (-0.5, -2.75) {};
		\node [style=none] (45) at (0.75, -2.75) {};
		\node [style=none] (46) at (0, 3.5) {};
		\node [style=none] (47) at (1.75, 0.75) {};
		\node [style=none] (48) at (-0.5, 1) {};
		\node [style=none] (49) at (1.75, -0.75) {};
		\node [style=small sg vertex] (50) at (0.75, -1.25) {d};
		\node [style=none] (51) at (0.75, 0.25) {};
		\node [style=none] (52) at (1.75, 0.25) {};
		\node [style=none] (53) at (3.5, 0) {$\rewritesto$};
		\node [style=none] (54) at (2.75, -1.25) {};
		\node [style=square box] (55) at (1.25, -0.25) {ripple};
		\node [style=square box] (56) at (0, 2) {induct};
		\node [style=none] (57) at (1.75, -1.25) {};
		\node [style=none] (58) at (0.5, 1) {};
		\node [style=none] (59) at (0, 2.5) {};
		\node [style=none] (60) at (0.5, 1.5) {};
		\node [style=small sg vertex] (61) at (1.75, -1.25) {e};
		\node [style=none] (62) at (-0.5, 1.5) {};
		\node [style=none] (63) at (0.75, -0.75) {};
		\node [style=none] (64) at (-0.5, 0.25) {};
		\node [style=none] (65) at (2.75, 0.75) {};
		\node [style=small sg vertex] (66) at (4.75, -1.25) {c};
		\node [style=none] (67) at (4.75, -2.75) {};
		\node [style=none] (68) at (6, -2.75) {};
		\node [style=none] (69) at (5.25, 3.5) {};
		\node [style=none] (70) at (7, 0.75) {};
		\node [style=none] (71) at (4.75, 1) {};
		\node [style=none] (72) at (7, -0.75) {};
		\node [style=small sg vertex] (73) at (6, -2) {d};
		\node [style=none] (74) at (6, 0.25) {};
		\node [style=none] (75) at (7, 0.25) {};
		\node [style=none] (76) at (8.75, 0) {$\rewritesto$};
		\node [style=none] (77) at (8, -1.25) {};
		\node [style=square box] (78) at (6.5, -0.25) {ripple};
		\node [style=square box] (79) at (5.25, 2) {induct};
		\node [style=none] (80) at (7, -1.25) {};
		\node [style=none] (81) at (5.75, 1) {};
		\node [style=none] (82) at (5.25, 2.5) {};
		\node [style=none] (83) at (5.75, 1.5) {};
		\node [style=small sg vertex] (84) at (7, 0.75) {e};
		\node [style=none] (85) at (4.75, 1.5) {};
		\node [style=none] (86) at (6, -0.75) {};
		\node [style=none] (87) at (4.75, 0.25) {};
		\node [style=none] (88) at (8, 0.75) {};
		\node [style=none] (89) at (13.25, 0.75) {};
		\node [style=none] (90) at (12.25, 0.25) {};
		\node [style=none] (91) at (11, 1.5) {};
		\node [style=none] (92) at (11.25, 0.25) {};
		\node [style=small sg vertex] (93) at (10, -1.25) {c};
		\node [style=small sg vertex] (94) at (11.25, -1.25) {f};
		\node [style=none] (95) at (10, 0.25) {};
		\node [style=none] (96) at (10.5, 2.5) {};
		\node [style=square box] (97) at (10.5, 2) {induct};
		\node [style=none] (98) at (10, 1) {};
		\node [style=none] (99) at (10, 1.5) {};
		\node [style=small sg vertex] (100) at (11.25, -2) {d};
		\node [style=none] (101) at (11, 1) {};
		\node [style=square box] (102) at (11.75, -0.25) {ripple};
		\node [style=none] (103) at (11.25, -0.75) {};
		\node [style=none] (104) at (12.25, -0.75) {};
		\node [style=none] (105) at (10, -2.75) {};
		\node [style=none] (106) at (11.25, -2.75) {};
		\node [style=none] (107) at (12.25, -1.25) {};
		\node [style=none] (108) at (10.5, 3.5) {};
		\node [style=none] (109) at (12.25, 0.75) {};
		\node [style=none] (110) at (13.25, -1.25) {};
		\node [style=small sg vertex] (111) at (10, -2) {b};
		\node [style=small sg vertex] (112) at (4.75, -2) {b};
		\node [style=small sg vertex] (113) at (-0.5, -2) {b};
		\node [style=small sg vertex] (114) at (-5.75, 1) {c};
	\end{pgfonlayer}
	\begin{pgfonlayer}{edgelayer}
		\draw [style=diredge] (4.center) to (10.center);
		\draw [style=diredge, in=90, out=-90, looseness=1.25] (3.center) to (5.center);
		\draw [style=diredge, in=90, out=-90] (12.center) to (7.center);
		\draw [in=90, out=-90, looseness=1.25] (11.center) to (8.center);
		\draw [style=diredge] (8.center) to (9.center);
		\draw [in=-90, out=-90, looseness=2.00] (1.center) to (15.center);
		\draw (15.center) to (13.center);
		\draw [in=90, out=90, looseness=2.00] (13.center) to (14.center);
		\draw (17.center) to (1.center);
		\draw (18.center) to (3.center);
		\draw (19.center) to (11.center);
		\draw [style=diredge] (14.center) to (20.center);
		\draw [style=diredge] (23.center) to (36.center);
		\draw [style=diredge, in=90, out=-90, looseness=1.25] (35.center) to (28.center);
		\draw [style=diredge, in=90, out=-90] (40.center) to (22.center);
		\draw [in=90, out=-90, looseness=1.25] (25.center) to (41.center);
		\draw [style=diredge] (41.center) to (21.center);
		\draw [in=-90, out=-90, looseness=2.00] (34.center) to (31.center);
		\draw (31.center) to (42.center);
		\draw [in=90, out=90, looseness=2.00] (42.center) to (24.center);
		\draw (26.center) to (34.center);
		\draw (37.center) to (35.center);
		\draw (39.center) to (25.center);
		\draw [style=diredge] (24.center) to (29.center);
		\draw [style=diredge] (46.center) to (59.center);
		\draw [style=diredge, in=90, out=-90, looseness=1.25] (58.center) to (51.center);
		\draw [style=diredge, in=90, out=-90] (63.center) to (45.center);
		\draw [in=90, out=-90, looseness=1.25] (48.center) to (64.center);
		\draw [style=diredge] (64.center) to (44.center);
		\draw [in=-90, out=-90, looseness=2.00] (57.center) to (54.center);
		\draw (54.center) to (65.center);
		\draw [in=90, out=90, looseness=2.00] (65.center) to (47.center);
		\draw (49.center) to (57.center);
		\draw (60.center) to (58.center);
		\draw (62.center) to (48.center);
		\draw [style=diredge] (47.center) to (52.center);
		\draw [style=diredge] (69.center) to (82.center);
		\draw [style=diredge, in=90, out=-90, looseness=1.25] (81.center) to (74.center);
		\draw [style=diredge, in=90, out=-90] (86.center) to (68.center);
		\draw [in=90, out=-90, looseness=1.25] (71.center) to (87.center);
		\draw [style=diredge] (87.center) to (67.center);
		\draw [in=-90, out=-90, looseness=2.00] (80.center) to (77.center);
		\draw (77.center) to (88.center);
		\draw [in=90, out=90, looseness=2.00] (88.center) to (70.center);
		\draw (72.center) to (80.center);
		\draw (83.center) to (81.center);
		\draw (85.center) to (71.center);
		\draw [style=diredge] (70.center) to (75.center);
		\draw [style=diredge] (108.center) to (96.center);
		\draw [style=diredge, in=90, out=-90, looseness=1.25] (101.center) to (92.center);
		\draw [style=diredge, in=90, out=-90] (103.center) to (106.center);
		\draw [in=90, out=-90, looseness=1.25] (98.center) to (95.center);
		\draw [style=diredge] (95.center) to (105.center);
		\draw [in=-90, out=-90, looseness=2.00] (107.center) to (110.center);
		\draw (110.center) to (89.center);
		\draw [in=90, out=90, looseness=2.00] (89.center) to (109.center);
		\draw (104.center) to (107.center);
		\draw (91.center) to (101.center);
		\draw (99.center) to (98.center);
		\draw [style=diredge] (109.center) to (90.center);
	\end{pgfonlayer}
\end{tikzpicture}}
\endpgfgraphicnamed}
  \end{center}
  Suppose applying induction to goal $a$ yields two base cases $b$, $c$ and a step case $d$. Then, in the first step, $a$ is consumed and $b,c$ are output on the first wire (of type \textit{base}) and $d$ is output on the second wire (of type \textit{step}). Then \textbf{ripple} is repeatedly applied until all sub-goals are on the output wires.
\end{example}

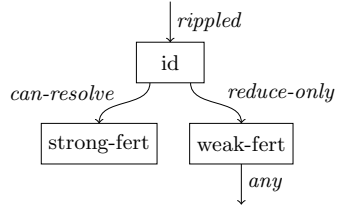
\begin{figure}
 \centering
   \vspace{-12pt}
 \scalebox{0.9}{%
\beginpgfgraphicnamed{fert_unfold_rhs}
\begin{tikzpicture}[string graph]
	\begin{pgfonlayer}{nodelayer}
		\node [style=none] (0) at (0, 2.5) {};
		\node [style=west wire label] (1) at (0.25, 2) {rippled};
		\node [style=square box, minimum width=1 cm] (2) at (0, 1) {id};
		\node [style=none] (3) at (0, 1.5) {};
		\node [style=none] (4) at (-0.5, 0.5) {};
		\node [style=east wire label] (5) at (-1.5, 0.25) {can-resolve};
		\node [style=none] (6) at (-1.75, -0.5) {};
		\node [style=square box] (7) at (-1.75, -1) {strong-fert};
		\node [style=none] (8) at (1.75, -0.5) {};
		\node [style=west wire label] (9) at (1.5, 0.25) {reduce-only};
		\node [style=none] (10) at (0.5, 0.5) {};
		\node [style=square box] (11) at (1.75, -1) {weak-fert};
		\node [style=west wire label] (12) at (2, -2) {any};
		\node [style=none] (13) at (1.75, -2.5) {};
		\node [style=none] (14) at (1.75, -1.5) {};
	\end{pgfonlayer}
	\begin{pgfonlayer}{edgelayer}
		\draw [style=diredge, in=90, out=-90] (0.center) to (3.center);
		\draw [style=diredge, in=90, out=-90] (4.center) to (6.center);
		\draw [style=diredge, in=90, out=-90] (10.center) to (8.center);
		\draw [style=diredge, in=90, out=-90] (14.center) to (13.center);
	\end{pgfonlayer}
\end{tikzpicture}}
\endpgfgraphicnamed}
 \vspace{-5pt}
 \caption{Fertilisation}\label{fig:fert}
   \vspace{-14pt}
\end{figure}

Proof strategies can easily become very large and complex. In PSGraph, we can
reduce this complexity and size by hiding parts of a graph -- achieved by
boxing a subgraphs into a single vertex. This box can be evaluated by
evaluating the graph it contains, or it may be unfolded in place. One
example of such hierarchy, is the `fertilise' box of Fig. \ref{fig:rippling},
which is shown in Fig. \ref{fig:fert}. Here, the `id' tactic simply returns
the input goal (e.g. \textsf{idtac} in Coq or \textsf{all\_tac} in Isabelle),
however it is used to route the input goal to the correct tactic, using the
goal types of the output wires. Here, we separate the case where the
goal can be resolved directly with the IH (called `strong fertilisation'),
from the case where the IH can only be used to reduce the goal (`weak fertilisation').
Note that the input and output wires of a nested graph must be the same as the
node which contains it. It is also possible in the PSGraph language to nest multiple graphs
in a single node, which can be used to produce branching OR/ORELSE behaviour, as detailed in
Section~\ref{sec:graph-tactics}.

\beforesection
\section{Goal Types}\label{sec:tactic}
\aftersection

For a type $\tau$, let $[\tau]$ be the type of finite lists and $\{ \tau \}$ be the type of finite sets whose elements are of type $\tau$.

Rather than considering all goals as members of one big type ``\textbf{goal}'', assume that we have a 
set of goal types $\mathcal G$. A particular goal type $\alpha \in \mathcal G$ represents all goals with some particular features, which may include local properties like ``contains symbol $X$'', proof state properties 
such as available facts, global properties like shared meta-variables, 
and relational properties with parent and possible children goals. Others have developed more detailed type theories for tactics (e.g. \cite{Stampoulis10}) which are closely related to our notion of a goal type. However, for our purposes it is sufficient to see a goal type as a predicate defined on goals:

\begin{definition}\label{def:goal-type} \rm
A \emph{goal type} $\alpha$ is a predicate $\alpha : \textbf{goal} \to \textbf{bool}$. Two goal types are said to be \emph{orthogonal}, written $\alpha \perp \beta$, if for all goals $g$, $\lnot\big(\alpha(g) \wedge \beta(g)\big)$.
\end{definition} 

The focus in this paper is on the use of goal types in the diagrammatic language, and the underlying theory is therefore beyond the scope of the paper. In fact, a PSGraph 
is generic w.r.t. the underlying goal type as it only relies on predicates as in Definition~\ref{def:goal-type}.
However, in order to illustrate goal types, we will use the following example of a goal type in the remainder of this paper:
\begin{example}
The following BNF shows the syntax of a goal type with a description of what it means:
\begin{center}
\begin{tabular}{rll}
\;\;GT := 
 & \textit{top\_symbol}($x_1,\cdots,x_n$) \!\desc{the top symbol of the goal is one of: $x_1,\cdots,x_n$} \\
$|$ & \textit{inductable} \!\desc{structural induction is applicable} \\
$|$ &  \textit{hyp\_embeds} \!\desc{hypothesis embeds in the goal} \\
$|$ & \textit{measure\_reducible} \!\desc{a measure towards a hypothesis is possible to reduce}\\
$|$ & \textit{hyp\_subst} $|$ \textit{hyp\_bck\_res} \!\desc{hypothesis applicable as rewrite/resolution rule}\\
$|$ & $GT_1 ~;~ GT_2$ $~|~$  $or(GT_1 ~...~ GT_N)$ \!\desc{conjunction and disjunction}\\
$|$ & $not(GT)$ $~|~$ \textit{any} \!\desc{negation and always succeed} \\
\end{tabular}
\end{center}
Whilst being relatively simple, $GT$ captures a range of properties, including all of the goal types
from Figs. \ref{fig:rippling} and \ref{fig:fert}:
\vspace{-2mm}
\begin{align*}
\textit{base} & = not(hyp\_embeds) \\
step = \textit{can\_ripple} & = hyp\_embeds;\,measure\_reduces \\
\textit{rippled} & = not(measure\_reducible); or(hyp\_bck\_res,hyp\_subst) \\
\textit{can\_resolve} & = \textit{hyp\_bck\_res};\textit{hyp\_embeds} \\
\textit{reduce\_only} & = \textit{not}(\textit{hyp\_bck\_res}); \textit{hyp\_subst}; \textit{hyp\_embeds}
\end{align*}
\end{example}
A richer goal type for the PSGraph framework, developed to support goal type generalisation for machine learning new graphs from 
example proofs, is defined in \cite{grov13a}.

The usual notion of a tactic can be treated as a function of the form:
\begin{equation}\label{eq:tac}
  \textbf{tac} : \textbf{goal} \to \{ [\textbf{goal}] \}
\end{equation}
That is, it takes a single goal to a set whose elements are lists of sub-goals. Each element of the set represents a branch in the (possibly non-deterministic) tactic evaluation. Note that we assume that 
internal details such as the production of an LCF justification function or direct modification of the proof state (a la Isabelle \cite{paper:Paulson:90}), are implicitly handled by the tactic. These details are not necessary to give the semantics of PSGraph evaluation, but shall play a role in the implementation of PSGraph in a particular prover, as discussed in Section~\ref{sec:impl}.

For a list $L$, we say a list of lists $L'$ is an \textit{ordered partition} if all of the lists are distinct, $L'$ contains the same elements as $L$ and each $l \in L'$ is obtained by deleting zero or more elements of $L$ (i.e. the order of $L$ is preserved).

\begin{definition} \rm
For goal types $\beta_1, \ldots, \beta_n$ and a list of goals $[g_1, \ldots, g_m]$, a \textit{type-partition} is an ordered partition:
\( P = [[g_i, g_{i'}, \ldots], [g_j, g_{j'}, \ldots], \ldots] \)
such that the $k$-th list in $P$ contains only goals of type $\beta_k$.
\end{definition}
In general, there may be more than one way to partition a list of goals. Let $\textbf{part}([\beta_1, \ldots, \beta_n], [g_1, \ldots, g_m])$ be the set of all possible partitions. The set of partitions is empty precisely when there is a goal in $L$ that is not of type $\beta_k$ for any $k$. Furthermore, if all of the goal types are orthogonal, this set must either be empty or a singleton.

\beforesection
\section{Evaluation of Proof Strategy Graphs}\label{sec:lang}
\aftersection

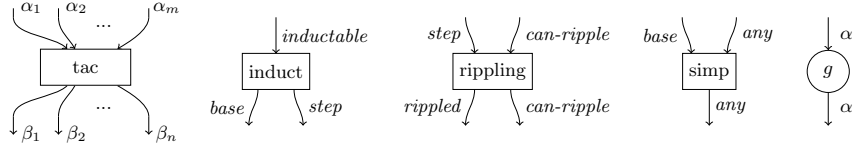
\begin{figure}
  \centering
  \scalebox{0.8}{%
\beginpgfgraphicnamed{tac_many_inputs}
\begin{tikzpicture}[string graph]
	\begin{pgfonlayer}{nodelayer}
		\node [style=none] (0) at (0.5, -1) {...};
		\node [style=square box, minimum width=1.5 cm, yshift=0.7 mm] (1) at (0, 0) {tac};
		\node [style=none] (2) at (-0.75, -1.25) {};
		\node [style=none] (3) at (-2, -1.25) {};
		\node [style=none] (4) at (1.75, -1.25) {};
		\node [style=none] (5) at (1.75, -1.75) {};
		\node [style=none] (6) at (-0.75, -1.75) {};
		\node [style=none] (7) at (-2, -1.75) {};
		\node [style=none, wire label] (8) at (-1.5, -1.75) {$\beta_1$};
		\node [style=none, wire label] (9) at (2.25, -1.75) {$\beta_n$};
		\node [style=none, wire label] (10) at (-0.25, -1.75) {$\beta_2$};
		\node [style=none] (11) at (1.75, 1.75) {};
		\node [style=none, wire label] (12) at (-0.25, 1.75) {$\alpha_2$};
		\node [style=none, wire label] (13) at (-1.5, 1.75) {$\alpha_1$};
		\node [style=none, wire label] (14) at (2.25, 1.75) {$\alpha_m$};
		\node [style=none] (15) at (-0.75, 1.5) {};
		\node [style=none] (16) at (-2, 1.75) {};
		\node [style=none] (17) at (0.5, 1.25) {...};
		\node [style=none] (18) at (1.75, 1.5) {};
		\node [style=none] (19) at (-2, 1.5) {};
		\node [style=none] (20) at (-0.75, 1.75) {};
	\end{pgfonlayer}
	\begin{pgfonlayer}{edgelayer}
		\draw [in=90, out=-135, looseness=0.75] (1) to (3.center);
		\draw [in=90, out=-120] (1) to (2.center);
		\draw [in=90, out=-30, looseness=0.75] (1) to (4.center);
		\draw [style=diredge] (3.center) to (7.center);
		\draw [style=diredge] (2.center) to (6.center);
		\draw [style=diredge] (4.center) to (5.center);
		\draw (16.center) to (19.center);
		\draw (20.center) to (15.center);
		\draw (11.center) to (18.center);
		\draw [style=diredge, in=135, out=-90, looseness=0.75] (19.center) to (1);
		\draw [style=diredge, in=120, out=-90] (15.center) to (1);
		\draw [style=diredge, in=30, out=-90, looseness=0.75] (18.center) to (1);
	\end{pgfonlayer}
\end{tikzpicture}}
\endpgfgraphicnamed\ \ %
\beginpgfgraphicnamed{tac_boxes}
\begin{tikzpicture}[string graph]
	\begin{pgfonlayer}{nodelayer}
		\node [style=west wire label] (0) at (-5.75, 1) {inductable};
		\node [style=none] (1) at (-6, 1.5) {};
		\node [style=square box] (2) at (-6, 0) {induct};
		\node [style=none] (3) at (-6, 0.5) {};
		\node [style=none] (4) at (-0.5, 0.5) {};
		\node [style=none] (5) at (-0.75, 1.5) {};
		\node [style=east wire label] (6) at (-1, 1) {step};
		\node [style=square box] (7) at (0, 0) {rippling};
		\node [style=east wire label] (8) at (-1, -1) {rippled};
		\node [style=none] (9) at (-5.25, -1.5) {};
		\node [style=none] (10) at (-6.5, -0.5) {};
		\node [style=east wire label] (11) at (-7, -1) {base};
		\node [style=none] (12) at (-6.75, -1.5) {};
		\node [style=west wire label] (13) at (-5, -1) {step};
		\node [style=none] (14) at (-5.5, -0.5) {};
		\node [style=none] (15) at (0.5, 0.5) {};
		\node [style=west wire label] (16) at (1, 1) {can-ripple};
		\node [style=none] (17) at (0.75, 1.5) {};
		\node [style=none] (18) at (0.75, -1.5) {};
		\node [style=none] (19) at (-0.5, -0.5) {};
		\node [style=none] (20) at (0.5, -0.5) {};
		\node [style=none] (21) at (-0.75, -1.5) {};
		\node [style=west wire label] (22) at (1, -1) {can-ripple};
		\node [style=none] (23) at (6, -1.5) {};
		\node [style=none] (24) at (5.5, 0.5) {};
		\node [style=square box] (25) at (6, 0) {simp};
		\node [style=none] (26) at (5.25, 1.5) {};
		\node [style=west wire label] (27) at (6.25, -1) {any};
		\node [style=none] (28) at (6, -0.5) {};
		\node [style=east wire label] (29) at (5, 1) {base};
		\node [style=west wire label] (30) at (7, 1) {any};
		\node [style=none] (31) at (6.75, 1.5) {};
		\node [style=none] (32) at (6.5, 0.5) {};
	\end{pgfonlayer}
	\begin{pgfonlayer}{edgelayer}
		\draw [style=diredge] (1.center) to (3.center);
		\draw [style=diredge, in=90, out=-90] (5.center) to (4.center);
		\draw [style=diredge, in=90, out=-90] (14.center) to (9.center);
		\draw [style=diredge, in=90, out=-90] (10.center) to (12.center);
		\draw [style=diredge, in=90, out=-90] (17.center) to (15.center);
		\draw [style=diredge, in=90, out=-90] (20.center) to (18.center);
		\draw [style=diredge, in=90, out=-90] (19.center) to (21.center);
		\draw [style=diredge, in=90, out=-90] (26.center) to (24.center);
		\draw [style=diredge, in=90, out=-90] (28.center) to (23.center);
		\draw [style=diredge, in=90, out=-90] (31.center) to (32.center);
	\end{pgfonlayer}
\end{tikzpicture}}
\endpgfgraphicnamed \ \  %
\beginpgfgraphicnamed{goal_node}
\begin{tikzpicture}[string graph]
	\begin{pgfonlayer}{nodelayer}
		\node [style=labelled sg vertex] (0) at (0, 0) {$g$};
		\node [style=none] (1) at (0, -1.5) {};
		\node [style=none] (2) at (0, 1.5) {};
		\node [style=none, wire label] (3) at (0.5, 1) {$\alpha$};
		\node [style=none, wire label] (4) at (0.5, -1) {$\alpha$};
	\end{pgfonlayer}
	\begin{pgfonlayer}{edgelayer}
		\draw [style=diredge] (2.center) to (0);
		\draw [style=diredge] (0) to (1.center);
	\end{pgfonlayer}
\end{tikzpicture}}
\endpgfgraphicnamed}
  \caption{Left to right: A generic tactic, 3 example tactics and a goal node}\label{fig:tac-and-goal-nodes}
\end{figure}

\noindent As already mentioned in section~\ref{sec:psgraph}, a PSGraph is a string diagram whose wires are labelled with goal types with two kinds of nodes: tactic nodes and goal nodes (Fig. \ref{fig:tac-and-goal-nodes}). Tactic nodes, represented as boxes, are labelled by the name of a tactic function of the form given in (\ref{eq:tac}) and have at least one input and zero or more outputs. A goal node is represented as a circle with exactly one input and output. 

Suppose a goal node $g$ occurs on an input wire of a tactic node labelled `tac', with output types $\beta_1, \ldots, \beta_n$. The goal node $g$ is propagated through the tactic node via a set of rewrite rules defined as follows:
\begin{enumerate}
  \item Evaluate $\textbf{tac}(g)$ to obtain a set of results (lists of sub-goals) from the tactic
  \item For each result $R \in \textbf{tac}(g)$ form a set of type-partitions: $\textbf{part}([\beta_1,\dots,\beta_n], R)$
  \item For each type-partition
  \( [[h_1, h_1', \ldots], \ldots, [h_n,h_n', \ldots]]
       \in \textbf{part}([\beta_1,\dots,\beta_n], R), \)
  define a rewrite rule where the input goal in the LHS is consumed in the RHS and each sub-goals of $[h_k,h_k',\ldots]$ are added to the $k$-th output wire of the RHS:
  \begin{equation}\label{eq:eval-rewrite}
    \scalebox{0.8}{%
\beginpgfgraphicnamed{goal_eval}
\begin{tikzpicture}[string graph]
	\begin{pgfonlayer}{nodelayer}
		\node [style=none, wire label] (0) at (-3.75, 3.75) {$\alpha$};
		\node [style=labelled sg vertex, inner sep=1 pt] (1) at (-4.25, 2.5) {$g$};
		\node [style=none] (2) at (-4.25, 3.75) {};
		\node [style=none] (3) at (-3.5, -1.75) {...};
		\node [style=none] (4) at (-0.5, 0) {$\rewritesto$};
		\node [style=none] (5) at (6, 0) {};
		\node [style=none] (6) at (2, 0) {};
		\node [style=none] (7) at (2, -4) {};
		\node [style=none] (8) at (4.75, -2.25) {...};
		\node [style=none, wire label] (9) at (6.5, -4) {$\beta_n$};
		\node [style=none, wire label] (10) at (2.5, -4) {$\beta_1$};
		\node [style=none] (11) at (4, 3.75) {};
		\node [style=none, wire label] (12) at (4.5, 3.75) {$\alpha$};
		\node [style=labelled sg vertex] (13) at (2, -1.5) {$h_1'$};
		\node [style=labelled sg vertex] (14) at (6, -1.5) {$h_n'$};
		\node [style=none] (15) at (6, -4) {};
		\node [style=none] (16) at (3.5, 0) {};
		\node [style=none, wire label] (17) at (4, -4) {$\beta_2$};
		\node [style=labelled sg vertex] (18) at (3.5, -1.5) {$h_2'$};
		\node [style=none] (19) at (3.5, -4) {};
		\node [style=square box, minimum width=1.5 cm, yshift=0.7 mm] (20) at (4, 1.5) {tac};
		\node [style=none] (21) at (2.5, 3) {...};
		\node [style=none] (22) at (6, 2.75) {};
		\node [style=none] (23) at (4.75, 3) {};
		\node [style=none] (24) at (2, 2.75) {};
		\node [style=none] (25) at (5.5, 3) {...};
		\node [style=none] (26) at (3.25, 3) {};
		\node [style=square box, minimum width=1.5 cm, yshift=0.7 mm] (27) at (-4.25, 0.5) {tac};
		\node [style=none] (28) at (-5, -1.5) {};
		\node [style=none] (29) at (-6.25, -1.5) {};
		\node [style=none] (30) at (-3.5, 2) {};
		\node [style=none] (31) at (-6.25, 1.75) {};
		\node [style=none] (32) at (-2.25, -1.5) {};
		\node [style=none] (33) at (-5.75, 2) {...};
		\node [style=none] (34) at (-5, 2) {};
		\node [style=none] (35) at (-2.25, 1.75) {};
		\node [style=none] (36) at (-2.75, 2) {...};
		\node [style=none] (37) at (-2.25, -2.75) {};
		\node [style=none] (38) at (-5, -2.75) {};
		\node [style=none] (39) at (-6.25, -2.75) {};
		\node [style=none, wire label] (40) at (-5.75, -2.75) {$\beta_1$};
		\node [style=none, wire label] (41) at (-1.75, -2.75) {$\beta_n$};
		\node [style=none, wire label] (42) at (-4.5, -2.75) {$\beta_2$};
		\node [style=labelled sg vertex] (43) at (3.5, -3) {$h_2$};
		\node [style=labelled sg vertex] (44) at (2, -3) {$h_1$};
		\node [style=labelled sg vertex] (45) at (6, -3) {$h_n$};
		\node [style=none] (46) at (6, -0.25) {...};
		\node [style=none] (47) at (2, -0.5) {};
		\node [style=none] (48) at (3.5, -0.25) {...};
		\node [style=none] (49) at (3.5, -0.5) {};
		\node [style=none] (50) at (6, -0.5) {};
		\node [style=none] (51) at (2, -0.25) {...};
	\end{pgfonlayer}
	\begin{pgfonlayer}{edgelayer}
		\draw [style=diredge] (2.center) to (1);
		\draw [style=diredge] (26.center) to (20);
		\draw [style=diredge, bend left=15] (24.center) to (20);
		\draw [style=diredge] (23.center) to (20);
		\draw [style=diredge, bend right=15] (22.center) to (20);
		\draw [style=diredge] (11.center) to (20);
		\draw [style=diredge, in=75, out=-139, looseness=0.75] (20) to (6.center);
		\draw [style=diredge, in=90, out=-106] (20) to (16.center);
		\draw [style=diredge, in=105, out=-41] (20) to (5.center);
		\draw [style=diredge] (34.center) to (27);
		\draw [style=diredge, bend left=15] (31.center) to (27);
		\draw [style=diredge] (30.center) to (27);
		\draw [style=diredge, bend right=15] (35.center) to (27);
		\draw [in=90, out=-139, looseness=0.75] (27) to (29.center);
		\draw [in=90, out=-105] (27) to (28.center);
		\draw [in=90, out=-41] (27) to (32.center);
		\draw [style=diredge] (1) to (27);
		\draw [style=diredge] (29.center) to (39.center);
		\draw [style=diredge] (28.center) to (38.center);
		\draw [style=diredge] (32.center) to (37.center);
		\draw [style=diredge] (47.center) to (13);
		\draw [style=diredge] (13) to (44);
		\draw [style=diredge] (44) to (7.center);
		\draw [style=diredge] (49.center) to (18);
		\draw [style=diredge] (18) to (43);
		\draw [style=diredge] (43) to (19.center);
		\draw [style=diredge] (50.center) to (14);
		\draw [style=diredge] (14) to (45);
		\draw [style=diredge] (45) to (15.center);
	\end{pgfonlayer}
\end{tikzpicture}}
\endpgfgraphicnamed}
  \end{equation}
\end{enumerate}
We shall call this set of rewrite rules $\textbf{RW}(\textbf{tac}, [\beta_1,\dots,\beta_n], g)$. If this set is empty, this corresponds to a failure. If it is a singleton, this corresponds to deterministic evaluation.

\begin{example} \label{ex:even2}
Suppose a goal $a := \textit{even}(2*n)$ occurs on an input wire of the \textbf{induct} tactic, which applies a two-step induction on the naturals (creating two base cases). To evaluate $a$, we first compute the ruleset $\textbf{RW}(\textbf{induct}, [base,step], a)$ by applying the tactic $\textbf{induct}(a)$. There is only one possible induction to perform, so the \textbf{induct} tactic returns a single list of sub-goals $\{[b,c,d]\}$, where
\[
  b = \textit{even}(2*0),\quad c = \textit{even}(2*1)\quad\textrm{and}\quad d = \textit{even}(2*n) \vdash \textit{even}(2*S(S(n))).
\]
\end{example}

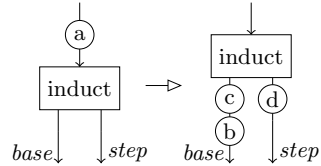
\begin{figure}
  \centering
  \vspace{-20pt}
  \scalebox{0.95}{%
\beginpgfgraphicnamed{rewrite_ex_ind}
\begin{tikzpicture}[string graph]
	\path [use as bounding box] (-3.75,-1.75) rectangle (3.25,2);
	\begin{pgfonlayer}{nodelayer}
		\node [style=small sg vertex] (0) at (-2.25, 1.25) {a};
		\node [style=none] (1) at (-1.75, -1) {};
		\node [style=none] (2) at (-2.25, 2) {};
		\node [style=none] (3) at (-1.75, -1.75) {};
		\node [style=square box] (4) at (-2.25, 0) {induct};
		\node [style=none] (5) at (-2.75, -1.75) {};
		\node [style=none] (6) at (-2.25, 0.5) {};
		\node [style=none] (7) at (-2.75, -1) {};
		\node [style=none] (8) at (-0.25, 0) {$\rewritesto$};
		\node [style=none] (9) at (-1.75, -0.5) {};
		\node [style=none] (10) at (-2.75, -0.5) {};
		\node [style=none] (11) at (1.75, 2) {};
		\node [style=none] (12) at (1.25, -0.75) {};
		\node [style=small sg vertex] (13) at (1.25, -1) {b};
		\node [style=none] (14) at (2.25, -1.75) {};
		\node [style=square box] (15) at (1.75, 0.75) {induct};
		\node [style=none] (16) at (2.25, -0.75) {};
		\node [style=none] (17) at (1.75, 1.25) {};
		\node [style=none] (18) at (2.25, 0.25) {};
		\node [style=small sg vertex] (19) at (2.25, -0.25) {d};
		\node [style=none] (20) at (1.25, 0.25) {};
		\node [style=none] (21) at (1.25, -1.75) {};
		\node [style=small sg vertex] (22) at (1.25, -0.25) {c};
		\node [style=west wire label] (23) at (-1.5, -1.5) {step};
		\node [style=east wire label] (24) at (-3, -1.5) {base};
		\node [style=west wire label] (25) at (2.5, -1.5) {step};
		\node [style=east wire label] (26) at (1, -1.5) {base};
	\end{pgfonlayer}
	\begin{pgfonlayer}{edgelayer}
		\draw [style=diredge] (2.center) to (6.center);
		\draw [style=diredge, in=90, out=-90, looseness=1.25] (1.center) to (3.center);
		\draw [style=diredge, in=90, out=-90, looseness=1.25] (7.center) to (5.center);
		\draw (9.center) to (1.center);
		\draw (10.center) to (7.center);
		\draw [style=diredge] (11.center) to (17.center);
		\draw [style=diredge, in=90, out=-90, looseness=1.25] (16.center) to (14.center);
		\draw [style=diredge, in=90, out=-90, looseness=1.25] (12.center) to (21.center);
		\draw (18.center) to (16.center);
		\draw (20.center) to (12.center);
	\end{pgfonlayer}
\end{tikzpicture}}
\endpgfgraphicnamed}
  \caption{Evaluation rule from Example \ref{ex:even2}}\label{fig:eval-ex}
  \vspace{-10pt}
\end{figure}

\noindent Next, the set of partitions $\textbf{part}([base,step],[b,c,d])$ is computed. Here, we see that $a$ and $b$ are \textit{base}
cases, as there are no hypothesis which can embed in the goal, while $d$ is a step case as the hypothesis does indeed embed 
in the goal. Thus, a single partition $[[b,c],[d]]$ is created. In the final step, the single rewrite rule (Fig.~\ref{fig:eval-ex}) is created. The result of applying this rule corresponds to the first step of example \ref{ex:even1}.

To evaluate a goal $g$ over a PSGraph $G$, we first add $g$ to an input of $G$ with a goal type which $g$ matches, then repeatedly apply rewrites generated by evaluating tactic nodes. By using PSGraph evaluation as a tactic in an LCF-style theorem prover, soundess will be guaranteed by the prover kernel. However, the next theorem states that evaluation is already ``as sound as the tactics it uses''.

\begin{theorem}[Soundness] \label{thm:soundness}
During PSGraph evaluation, goal nodes are only produced/consumed by calls to tactics, and never duplicated or lost during evaluation.
\end{theorem}

\begin{proof}
  Every rewrite rule applied during evaluation is the result of a call to the partition function $\textbf{part}$ on the output of a tactic, which yields rewrite rules where the input of a tactic is consumed and sub-goals produced by the tactic must each occur on precisely one output wire.
\end{proof}

\begin{definition} \rm
A PSGraph is said to be in \emph{terminal form} if the only goal nodes it contains are on output wires. 
\end{definition}

\begin{definition} \rm
Let $\mathcal T$ be a tree whose leaves are labelled with PSGraphs or $\bot$. Graph leaves in terminal form in $\mathcal T$ are said to be \textit{closed}. Otherwise, they are called \textit{open}. An \emph{evaluation strategy} is a function $S : \mathcal T \to \mathcal T$ which chooses an open PSGraph $G$ in $\mathcal T$ and unfolds it by: (i) selecting a goal $g$ on the input wire of a tactic node $\textbf{tac}$ and (ii) adding
the children arising from applying each of the rules $r \in \textbf{RW}(\textbf{tac}, [\beta_1,\dots,\beta_n], g)$, or a single child $\bot$ indicating failure, to $G$ in $\mathcal T$. We say $\mathcal{T}$ is \emph{terminated} when all graph leaves are closed.
\end{definition}

\begin{example}
A depth-first strategy $S_{DF}$ will select the open PSGraph that was last produced, and within it unfold the
goal that was last produced. A more sophisticated strategy $S_{S}$ may for example select the open PSGraph with the
fewest goals and evaluate the goal which is most likely to fail to cut a failed branch as early as possible.
\end{example}

\beforesection
\section{Combinators and Hierarchies}\label{sec:graph-tactics}
\aftersection

\begin{figure}
  \centering
  \scalebox{0.78}{%
\beginpgfgraphicnamed{combinators}
\begin{tikzpicture}[string graph]
	\begin{pgfonlayer}{nodelayer}
		\node [style=none] (0) at (-6.25, 2.5) {...};
		\node [style=none] (1) at (-5.25, 2.75) {};
		\node [style=cloud vertex, inner sep=2 mm, shape=cloud, aspect=1.5] (2) at (-6.25, 1.25) {$G$};
		\node [style=none] (3) at (-7.25, 2.75) {};
		\node [style=none] (4) at (-6.25, 0) {...};
		\node [style=none] (5) at (-5.25, -2.75) {};
		\node [style=none] (6) at (-6.25, -2.5) {...};
		\node [style=none] (7) at (-7.25, -2.75) {};
		\node [style=cloud vertex, inner sep=2 mm, shape=cloud, aspect=1.5] (8) at (-6.25, -1.25) {$H$};
		\node [style=none] (9) at (-6.25, 3.5) {$G \textrm{ THEN } H$};
		\node [style=east wire label] (10) at (-7.5, -2.75) {$\gamma_1$};
		\node [style=west wire label] (11) at (-5, -2.75) {$\gamma_n$};
		\node [style=west wire label] (12) at (-5, 2.75) {$\alpha_l$};
		\node [style=east wire label] (13) at (-7.5, 2.75) {$\alpha_1$};
		\node [style=west wire label] (14) at (-5, 0) {$\beta_m$};
		\node [style=east wire label] (15) at (-7.5, 0) {$\beta_1$};
		\node [style=cloud vertex, inner sep=2 mm, shape=cloud, aspect=1.5] (16) at (-1.5, 0) {$G$};
		\node [style=none] (17) at (-1.5, 1.25) {...};
		\node [style=none] (18) at (-0.5, -1.5) {};
		\node [style=cloud vertex, inner sep=2 mm, shape=cloud, aspect=1.5] (19) at (2.75, 0) {$G'$};
		\node [style=none] (20) at (-2.5, 1.5) {};
		\node [style=none] (21) at (0.75, 2.25) {$G \textrm{ TENSOR } G'$};
		\node [style=none] (22) at (2.75, 1.25) {...};
		\node [style=none] (23) at (1.75, -1.5) {};
		\node [style=none] (24) at (-0.5, 1.5) {};
		\node [style=none] (25) at (-2.5, -1.5) {};
		\node [style=none] (26) at (3.75, -1.5) {};
		\node [style=none] (27) at (-1.5, -1.25) {...};
		\node [style=none] (28) at (1.75, 1.5) {};
		\node [style=none] (29) at (3.75, 1.5) {};
		\node [style=none] (30) at (2.75, -1.25) {...};
		\node [style=west wire label] (31) at (-0.25, -1.5) {$\beta_m$};
		\node [style=east wire label] (32) at (-2.75, -1.5) {$\beta_1$};
		\node [style=west wire label] (33) at (-0.25, 1.5) {$\alpha_l$};
		\node [style=east wire label] (34) at (-2.75, 1.5) {$\alpha_1$};
		\node [style=west wire label] (35) at (4, 1.5) {$\alpha_l'$};
		\node [style=east wire label] (36) at (1.5, 1.5) {$\alpha_1'$};
		\node [style=west wire label] (37) at (4, -1.5) {$\beta_m'$};
		\node [style=east wire label] (38) at (1.5, -1.5) {$\beta_1'$};
		\node [style=none] (39) at (-4, -3.25) {};
		\node [style=none] (40) at (-8.25, -3.25) {};
		\node [style=none] (41) at (-8.25, 4) {};
		\node [style=none] (42) at (-4, 4) {};
		\node [style=none] (43) at (4.75, -2) {};
		\node [style=none] (44) at (4.75, 2.75) {};
		\node [style=none] (45) at (-3.5, 2.75) {};
		\node [style=none] (46) at (-3.5, -2) {};
		\node [style=cloud vertex, inner sep=2 mm, shape=cloud, aspect=1.5] (47) at (7.25, 0) {$G$};
		\node [style=none] (48) at (8.75, 1.25) {};
		\node [style=none] (49) at (7.25, 1.5) {...};
		\node [style=west wire label] (50) at (10, 0) {$\alpha$};
		\node [style=none] (51) at (6.25, 1.5) {};
		\node [style=none] (52) at (7.25, -1.5) {...};
		\node [style=none] (53) at (8.75, -1.25) {};
		\node [style=none] (54) at (6.25, -1.5) {};
		\node [style=east wire label] (55) at (6, 1.5) {$\gamma_1$};
		\node [style=east wire label] (56) at (6, -1.5) {$\beta_1$};
		\node [style=none] (57) at (8, 2.25) {$\textrm{REPEAT}_\alpha(G)$};
		\node [style=none] (58) at (10.5, 2.75) {};
		\node [style=none] (59) at (5.25, 2.75) {};
		\node [style=none] (60) at (5.25, -2) {};
		\node [style=none] (61) at (10.5, -2) {};
		\node [style=small square box, minimum width=1 cm] (62) at (15.25, -2) {id};
		\node [style=none] (63) at (14, 2) {...};
		\node [style=east wire label] (64) at (12.5, -2.75) {$\beta_1$};
		\node [style=cloud vertex, inner sep=2 mm, shape=cloud, aspect=1.5] (65) at (15.5, 0) {$H$};
		\node [style=none] (66) at (15.25, -3) {};
		\node [style=west wire label] (67) at (15.5, 2.75) {$\alpha_m$};
		\node [style=none] (68) at (15.25, 1.25) {...};
		\node [style=cloud vertex, inner sep=2 mm, shape=cloud, aspect=1.5] (69) at (12.5, 0) {$G$};
		\node [style=none] (70) at (15.25, -1.25) {...};
		\node [style=none] (71) at (14, -2) {...};
		\node [style=none] (72) at (12.75, -1.25) {...};
		\node [style=none] (73) at (15.25, 3) {};
		\node [style=small square box, minimum width=1 cm] (74) at (12.75, 2) {id};
		\node [style=none] (75) at (12.75, 3) {};
		\node [style=small square box, minimum width=1 cm] (76) at (12.75, -2) {id};
		\node [style=west wire label] (77) at (15.5, -2.75) {$\beta_n$};
		\node [style=none] (78) at (12.75, -3) {};
		\node [style=none] (79) at (12.75, 1.25) {...};
		\node [style=east wire label] (80) at (12.5, 2.75) {$\alpha_1$};
		\node [style=small square box] (81) at (15.25, 2) {id};
		\node [style=none] (82) at (17, 4) {};
		\node [style=none] (83) at (14, 3.5) {$G \textrm{ OR' } H$};
		\node [style=none] (84) at (11, 4) {};
		\node [style=none] (85) at (11, -3.25) {};
		\node [style=none] (86) at (17, -3.25) {};
	\end{pgfonlayer}
	\begin{pgfonlayer}{edgelayer}
		\draw [style=diredge, in=120, out=-90] (3.center) to (2);
		\draw [style=diredge, in=60, out=-90] (1.center) to (2);
		\draw [style=diredge, in=90, out=-124] (8) to (7.center);
		\draw [style=diredge, in=90, out=-56] (8) to (5.center);
		\draw [style=diredge, bend right=45] (2) to (8);
		\draw [style=diredge, bend left=45] (2) to (8);
		\draw [style=diredge, in=120, out=-90] (20.center) to (16);
		\draw [style=diredge, in=60, out=-90] (24.center) to (16);
		\draw [style=diredge, in=90, out=-60] (16) to (18.center);
		\draw [style=diredge, in=90, out=-120] (16) to (25.center);
		\draw [style=diredge, in=120, out=-90] (28.center) to (19);
		\draw [style=diredge, in=60, out=-90] (29.center) to (19);
		\draw [style=diredge, in=90, out=-60] (19) to (26.center);
		\draw [style=diredge, in=90, out=-120] (19) to (23.center);
		\draw [style=gray edge] (41.center) to (42.center);
		\draw [style=gray edge] (42.center) to (39.center);
		\draw [style=gray edge] (39.center) to (40.center);
		\draw [style=gray edge] (40.center) to (41.center);
		\draw [style=gray edge] (45.center) to (44.center);
		\draw [style=gray edge] (44.center) to (43.center);
		\draw [style=gray edge] (43.center) to (46.center);
		\draw [style=gray edge] (46.center) to (45.center);
		\draw [style=diredge, in=120, out=-90] (51.center) to (47);
		\draw [style=diredge, in=90, out=-120] (47) to (54.center);
		\draw [in=0, out=0, looseness=1.25] (53.center) to (48.center);
		\draw [style=diredge, in=45, out=180, looseness=0.75] (48.center) to (47);
		\draw [in=180, out=-45, looseness=0.75] (47) to (53.center);
		\draw [style=gray edge] (59.center) to (58.center);
		\draw [style=gray edge] (58.center) to (61.center);
		\draw [style=gray edge] (61.center) to (60.center);
		\draw [style=gray edge] (60.center) to (59.center);
		\draw [style=diredge, in=45, out=-135] (81) to (69);
		\draw [style=diredge, in=135, out=-135, looseness=1.25] (74) to (69);
		\draw [style=diredge] (75.center) to (74);
		\draw [style=diredge] (73.center) to (81);
		\draw [style=diredge, in=135, out=-45] (74) to (65);
		\draw [style=diredge, in=45, out=-45, looseness=1.25] (81) to (65);
		\draw [style=diredge] (76) to (78.center);
		\draw [style=diredge] (62) to (66.center);
		\draw [style=diredge, in=135, out=-135, looseness=1.25] (69) to (76);
		\draw [style=diredge, in=45, out=-45, looseness=1.25] (65) to (62);
		\draw [style=diredge, in=45, out=-135] (65) to (76);
		\draw [style=diredge, in=135, out=-45] (69) to (62);
		\draw [style=gray edge] (84.center) to (82.center);
		\draw [style=gray edge] (82.center) to (86.center);
		\draw [style=gray edge] (86.center) to (85.center);
		\draw [style=gray edge] (85.center) to (84.center);
	\end{pgfonlayer}
\end{tikzpicture}}
\endpgfgraphicnamed}
  \caption{\label{fig:then-and-tensor} THEN, TENSOR, $\textrm{REPEAT}_\alpha$, and OR' combinators}
\end{figure}
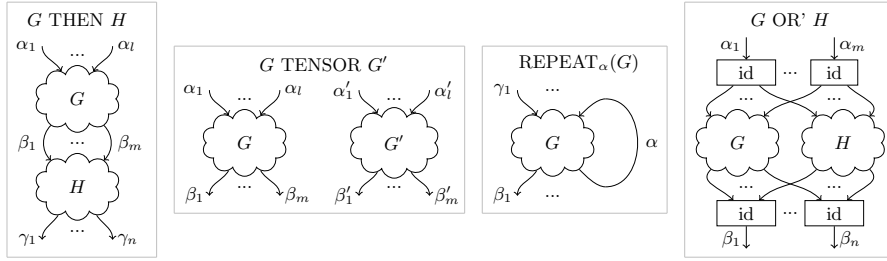

\noindent An interesting feature of graphical  languages is that it gives us many techniques for
combining strategies. In this section, we will discuss two such techniques: graph \textit{combinators} and graph \textit{hierarchies}.

Graph combinators can be used to syntactically build new strategy graphs from old graphs. Perhaps the simplest graph combinators are the THEN and TENSOR\footnote{We use TENSOR for parallel composition as this is common for graphical languages (see e.g. \cite{paper:Selinger:09}),
and has also been used in tactic languages such as HiTac  \cite{paper:Aspinall:2008}.}
combinators  (Fig. \ref{fig:then-and-tensor}). THEN takes all the outputs of one graph and connects them to all of the inputs of another graph\footnote{This process of plugging one or more inputs and outputs together is defined formally using graph pushouts in \cite{paper:Dixon:10}.}. TENSOR is at the other extreme: it combines two graphs into one without plugging any wires together.
The THEN combinator uses goal types on the wires to figure out which output should be connected to which input, i.e. an output of type $\beta_i$ in $G$ is always connected to an input of type $\beta_i$ in $H$. As a consequence, ``$G$ THEN $H$'' is only well-defined when the output types of $G$ match the input types of $H$ and all of the $\beta_i$ are distinct.

TENSOR can be thought of as a sort of ``parallel composition'' of strategies. In an expression like ``$G$ THEN ($H$ TENSOR $H'$)'', $H$ will handle some of the goals produced by $G$ and $H'$ will handle the rest. Which goal goes where is determined by the goal type.

 One could imagine many variations on the THEN combinator that perform various more general kinds of wire-pluggings, however, for space reasons, we consider just one more kind of plugging combinator called $\textrm{REPEAT}_\alpha$
(Fig. \ref{fig:then-and-tensor}). It  connects an output of type $\alpha$ to an input of type $\alpha$, introducing a feedback loop. As with the THEN combinator, $\textrm{REPEAT}_\alpha$ is not always well-defined. It is defined whenever the graph $G$ has precisely one input and one output of type $\alpha$. This is not much of a restriction, as input and output types of PSGraphs should typically be distinct to make the most of the goal typing system. Note also that   $\textrm{REPEAT}_\alpha$ is close in character to the traditional REPEAT\_WHILE tactical,
taking $\alpha$ to be the predicate controlling the repeated application.

Branching can be achieved by exploiting non-determinism of tactic node evaluation when faced with non-orthogonal output goal types.
This can be seen by the OR' combinator in Fig. \ref{fig:then-and-tensor}, which is a graphical variant of the OR combinator. 
However, when considering $G$ and $H$ as two distinct alternatives, each graph should really be considered in isolation, but this information is effectively lost by combining them into the same graph. For instance, there is nothing to stop us from adding a wire between them or interleaving evaluation of the two branches. Moreover, we cannot represent other more controlled types of branching, such as an ORELSE combinator.

In Section~\ref{sec:psgraph}, we saw that we can hide complexities by folding subgraphs into a single node in the graph. 
This was illustrated by the `fertilise' node for rippling. We call such a hierarchical node in a PSGraph a \emph{graph tactic}.
In addition to hiding complexity, a graph tactic can handle branching in a natural way, and allows us to mark specific subgraphs with different evaluation strategies. 

\begin{definition}\label{def:graph-tactic} \rm
  A \textit{graph tactic} $N$ contains a pair $(A, \mathcal G)$, consisting of a label $A \in \{ \textrm{OR}, \textrm{ORELSE} \}$ and a non-empty list of pairs $\mathcal G = [(G_1, S_1), \ldots, (G_n, S_n)]$, where all of the graphs $G_i$ have the same number and type of inputs/outputs as $N$ and each $S_i$ is an optional evaluation strategy for the graph $G_i$.  A tactic node that is not a graph tactic is called an \emph{atomic tactic}.
\end{definition}

For a graph tactic containing $\big(\textrm{OR},[(G_1,S_1),(G_2,S_2)]\big)$, we often omit the evaluation strategy and label this node $\textrm{OR}[G_1,G_2]$. In other cases we give the node an explicit name, as in e.g. `fertilise'.
The list $\mathcal G$ holds the graphs that are nested, and multiple elements in the list correspond to alternation.
The label OR/ORELSE is called the \textit{alternation style} of the graph tactic, and the OR and ORELSE combinators
can be naturally expressed with these alternation styles. OR is a branching search, attempting to evaluate each graph $G_i$ in turn. On the other hand, ORELSE proceeds sequentially until a \textit{single} graph is evaluated successfully. If
 $\mathcal G$ is a singleton list then the alternation style will have no impact on evaluation.
 
 \beforesection
\subsection{Evaluation \& Unfolding of Hierarchies}
\aftersection

So, it only remains to describe the evaluation of a single element $(G_i,S_i)$ of $\mathcal G$ of graph tactic `\textit{tac}'. 
This is achieved in the same
way as in Section~\ref{sec:lang}, by generating a set of evaluation rewrite rules. It deviates from evaluation
of such atomic tactics by the way the output nodes are generated. Let $L$ be the LHS of the usual evaluation rewrite rule (\ref{eq:eval-rewrite}), with goal node $g$ be on the $j$-th input wire of `\textit{tac}'. The set of evaluation rules from $(S_i,G_i)$ is then created as follows:
\begin{enumerate}
  \item Place $g$ on the $j$-th input wire of the graph $G_i$, which becomes the root of the singleton search tree $\mathcal{T}$.
  \item Let $S$ be $S_i$ if it is defined, if not let it be the evaluation strategy of the parent graph. Use $S$ to 
  evaluate $\mathcal{T}$ until $\mathcal{T}$ has terminated. 
  \item For each terminal leaf $G_i'$ of $\mathcal{T}$, there will be zero or more goals on each of the output wires. Let $R$ be
  $L$ with node $g$ removed.  For all $k$, place all of the goals on the $k$-th output wire of $G_i'$ on to the $k$-th output wire of \textit{tac} in $R$, in the same order. This yields a rewrite rule $L \rewritesto R$.
\end{enumerate}
Thus, there will be one rewrite rule for each terminal PSGraph in $\mathcal{T}$. 
This hierarchical evaluation procedure buys us two things at once. The first is modularity: complex strategies can be broken into multiple graph tactics composed in a high-level strategy graph. The second is fine-grained control over evaluation strategies: different subgraphs can be associated with different evaluation strategies, which can be tailored to the specific task at hand.

\begin{figure}
  \centering
  \vspace{-12pt}
  \scalebox{0.85}{%
\beginpgfgraphicnamed{unfold_OR}
\begin{tikzpicture}[string graph]
	\begin{pgfonlayer}{nodelayer}
		\node [style=cloud vertex, inner sep=2 mm, shape=cloud, aspect=1.5] (0) at (2.5, 0) {$G_i$};
		\node [style=none] (1) at (2.5, 1.5) {...};
		\node [style=none] (2) at (3.5, -1.75) {};
		\node [style=none] (3) at (1.5, 1.75) {};
		\node [style=none] (4) at (3.5, 1.75) {};
		\node [style=none] (5) at (1.5, -1.75) {};
		\node [style=none] (6) at (2.5, -1.5) {...};
		\node [style=west wire label] (7) at (3.75, -1.75) {$\beta_n$};
		\node [style=east wire label] (8) at (1.25, -1.75) {$\beta_1$};
		\node [style=west wire label] (9) at (3.75, 1.75) {$\alpha_m$};
		\node [style=east wire label] (10) at (1.25, 1.75) {$\alpha_1$};
		\node [style=none] (11) at (0, 0) {$\rewritesto$};
		\node [style=square box] (12) at (-3.25, 0) {$\textrm{OR}[G_1,\ldots,G_n]$};
		\node [style=west wire label] (13) at (-1.75, 1.75) {$\alpha_m$};
		\node [style=none] (14) at (-3.25, 1.5) {...};
		\node [style=none] (15) at (-4.5, 1.75) {};
		\node [style=none] (16) at (-2, 1.75) {};
		\node [style=east wire label] (17) at (-4.75, 1.75) {$\alpha_1$};
		\node [style=east wire label] (18) at (-4.75, -1.75) {$\beta_1$};
		\node [style=none] (19) at (-2, -1.75) {};
		\node [style=none] (20) at (-4.5, -1.75) {};
		\node [style=west wire label] (21) at (-1.75, -1.75) {$\beta_n$};
		\node [style=none] (22) at (-3.25, -1.5) {...};
		\node [style=none] (23) at (-2.5, 0.5) {};
		\node [style=none] (24) at (-4, 0.5) {};
		\node [style=none] (25) at (-2.5, -0.5) {};
		\node [style=none] (26) at (-4, -0.5) {};
	\end{pgfonlayer}
	\begin{pgfonlayer}{edgelayer}
		\draw [style=diredge, in=120, out=-90] (3.center) to (0);
		\draw [style=diredge, in=60, out=-90] (4.center) to (0);
		\draw [style=diredge, in=90, out=-60] (0) to (2.center);
		\draw [style=diredge, in=90, out=-120] (0) to (5.center);
		\draw [style=diredge, in=90, out=-90] (15.center) to (24.center);
		\draw [style=diredge, in=90, out=-90] (16.center) to (23.center);
		\draw [style=diredge, in=90, out=-90] (26.center) to (20.center);
		\draw [style=diredge, in=90, out=-90] (25.center) to (19.center);
	\end{pgfonlayer}
\end{tikzpicture}}
\endpgfgraphicnamed}
  \vspace{-2pt}
  \caption{An ``unfolding'' rule}\label{fig:unfold-or}
  \vspace{-14pt}
\end{figure}
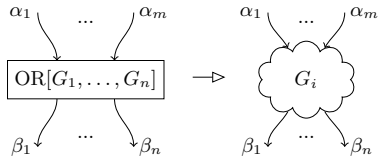

It is also worth noting that there is a second, rewriting-based method of expressing this hierarchical evaluation procedure. Since the graphs $G_1, \ldots, G_n$ in a graph tactic node have the same inputs and outputs as the node itself, we can define a rewrite rule for each $G_i$ (Fig.~\ref{fig:unfold-or}). This rule (and its inverse) give us a way to selectively unfold and re-fold parts of the graph. These rules can be used during evaluation to perform an \textit{in situ} version of the hierarchical evaluation procedure described above. Perhaps more interestingly, inspired by~\cite{paper:Whiteside:11}, they can be used during proof strategy design to refactor a complex strategy graph.

\beforesection
\section{Implementation}\label{sec:impl}
\aftersection

\newcommand{\psgraphtool}{\textsf{PSGraph}}

The PSGraph language is independent of both the underlying theorem prover and the goal types used. This is reflected in  our implementation, called \psgraphtool{}.\footnote{The tool is available at \url{https://github.com/ggrov/psgraph/tree/lpar13}.} It is implemented in
Poly/ML and consists of $4$ layers:
\begin{enumerate}
\item At the bottom is the core of the existing \emph{Quantomatic} graph rewriting system~\cite{Quantomatic}, which implements the (string diagram) theory from \cite{paper:Dixon:10}.
\item Then there is the \emph{generic PSGraph language layer}, which implements the features
described in Sections \ref{sec:tactic}$-$\ref{sec:graph-tactics} using Quantomatic.
\item On top of the PSGraph layer, there is the \emph{goal type layer}, where a goal node (wrapping a theorem proving specific sub-goal), a goal type and a matching function between them is defined. The generic layer is then instantiated with these features.
\item At the top is the \emph{theorem prover specific layer}, which instantiates the generic and goal type layers with theorem proving specific features. These include: the underlying proof and tactic representations, term/goal matching functions, and a set of tactics provided by the prover.
\end{enumerate}
The implementation discussed here contains an instantiation of the goal type $GT$ of Section \ref{sec:tactic} for Isabelle/HOL \cite{paper:Paulson:90}. The goal type in \cite{grov13a} and limited support for
the ProofPower theorem prover\footnote{See \url{http://www.lemma-one.com/ProofPower/index/}.} has also been implemented (also available from the \psgraphtool{} webpage).

\beforesection
\subsection{Proof Representation in Isabelle}
\aftersection

Theorem provers typically work by applying a tactic to one of the open sub-goals,
which either discharges the sub-goal, or generates new sub-goal which then has to
be discharged. This is repeated until there are no more sub-goals. The results of these applications must then be combined to create the actual proof. This step is handled
differently between provers: Isabelle combines these steps by having just one goal in which all the remaining ``sub-goals'' occur as premises, whereas HOL/ProofPower generates a ``justification function'' to combine sub-goals. Others, such as \cite{paper:Aspinall:2008,paper:Whiteside:11,Stampoulis10},
have given formal semantics to the relationship between tactics and 
the proofs produced. In the context of PSGraph, we see this as a theorem prover
specific task, and instead only focus on working with the open sub-goals produced. 
This is reflected by the fact that our key soundness property is the goal property highlighted in 
Theorem \ref{thm:soundness}. As a result, the proof representation has to be handled by the top layer in our architecture,
which instantiates the system for a particular prover.

To prove $F$ in Isabelle, the initial goal (henceforth proof) 
$F \Longrightarrow F$ is 
created, where $\Longrightarrow$ should be read as logical entailment.
If a tactic reduces $F$ to the sub-goals $G$ and $H$, then the proof  
becomes $G \Longrightarrow H \Longrightarrow F$. A tactic in Isabelle (normally) works on a particular sub-goal, and the index of this sub-goal must be provided. This will produce a set (lazy sequence to be exact) of new proofs, where each element is a branch. For example, let \textbf{tac} be a tactic which reduces $H$ to sub-goals $I$ and $J$. Then `\textbf{tac} $2$' applied to the above proof will give the (singleton) proof $G \Longrightarrow I \Longrightarrow J \Longrightarrow F$. When there are no sub-goals,
and we are left with just $F$, then the proof is completed. 

To handle this ``side effect" a tactic has on the proof object, during evaluation we keep track of an Isabelle proof $\textit{prf}$, paired with a map $m$ from a name to a sub-goal index. Then, for a goal $g$ and a tactic \textbf{tac},
the first step in the evaluation of Section \ref{sec:tactic} becomes:
\begin{itemize}
\item Look up the name of $g$ in $m$ to give the index $i$.
\item Apply $\textbf{tac}~i~\textit{prf}$, which creates a set of new proofs.
\item For each new proof: find the new sub-goals starting at position $i$; update all indices in $m$ to reflect the new sub-goals (e.g. if two sub-goals are created then all indices after $i$ have to be incremented by $1$); create a fresh name for each new-sub-goal and update $m$, and return the new sub-goals with their name.
\end{itemize}

\beforesection
\subsection{Isabelle/Isar Proof Method \& GUI}
\aftersection

\begin{figure}
\centering
\includegraphics[width=\textwidth]{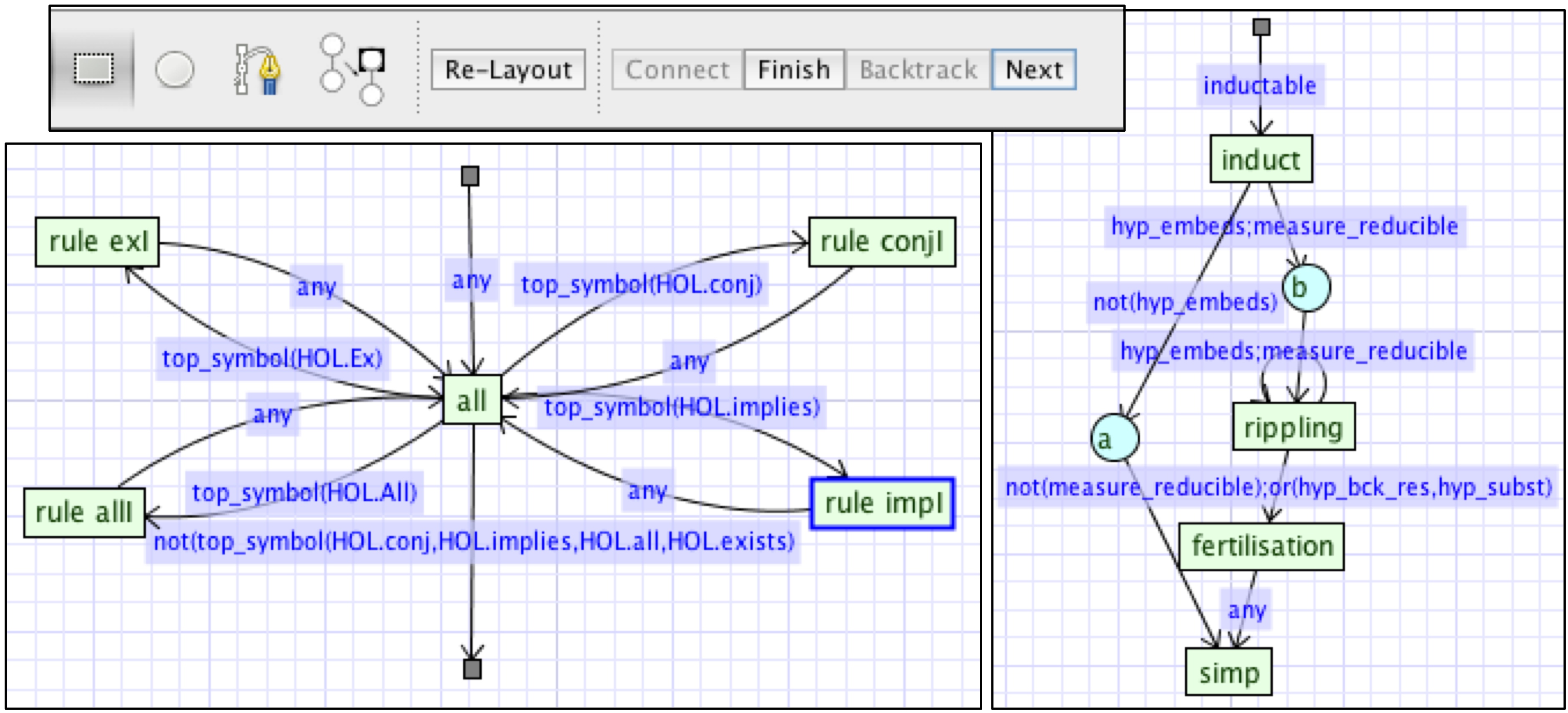}
\caption{GUI:  navigation bar (top), intro graph (left), rippling evaluation (right).} \label{fig:gui}
\vspace{-5mm}
\end{figure}

\psgraphtool{} has a GUI where users can both draw and, for a given conjecture, 
inspect the evaluation of a PSGraph.  Fig. \ref{fig:gui} shows some screen-shots 
of this GUI, which we will return to below. 

Our Isabelle instantiation is encoded as a new theory on top of the 
`Main' Isabelle/HOL theory\footnote{See \url{https://isabelle.in.tum.de/} for details.}. On top of this we have created a new proof method for Isabelle/Isar called
\textsf{psgraph} in order to make usage more Isabelle friendly. Graphs that
have been drawn, or implemented (using the combinators), must be 
explicitly registered in Isabelle with a name in order to use them.
They can then be used by the following Isabelle method in the middle of a proof:
\begin{center}\sf
\textbf{apply} (psgraph [(interactive)] $\langle$graph-name$\rangle$ [searchf: $\langle$sname$\rangle$]  [goalf: $\langle$ename$\rangle$])
\end{center}
\textsf{$\langle$graph-name$\rangle$} refers to the name of a registered graph.
The optional \textsf{(interactive)} flag enters a `debugging mode' where the user can use the GUI to step through a proof. The navigation bar in Fig. \ref{fig:gui} illustrates how the user can `Connect' to Isabelle, and step through (`Next') the proof. `Finish' will return to Isabelle, and all remaining sub-goals 
become sub-goals in the Isabelle proof. The evaluation strategies can be configured by \textsf{searchf}, with a name of a search strategy, and \textsf{goalf}, which selects
which goal to pick first. Finally, note that there is a special `current' mode for the interactive version, where the graph which is currently open in GUI is used. This option is selected by  `\textsf{\textbf{apply} (psgraph (current))}', and is useful for testing while strategies are being drawn. 

\vspace{10pt}

\noindent \textbf{Examples and Tool Evaluation.}
We have implemented the \emph{rippling} strategy in \psgraphtool{} as an 
adaptation of the version found in IsaPlanner \cite{paper:Dixon:03}. The right
hand side of Fig. \ref{fig:gui}, illustrates a rippling proof in interactive mode
with two open goals ($a$ and $b$). We have evaluated our rippling implementation
on $35$ Peano arithmetic and list examples. These can be seen and tested
by downloading the tool\footnote{See \url{https://github.com/ggrov/psgraph/tree/lpar13/src/examples/LPAR13}}. 
The butterfly-shaped strategy on the left of Fig. \ref{fig:gui} is an implementation of the well-known \emph{intro}-tactic as a PSGraph. This strategy supports `any' input goal, and uses $top\_symbol$, $any$ and $not$ $GT$ predicates. 
The $all$ node uses \textsf{all\_tac}, which is Isabelle's version of `\textit{id}', i.e. the tactic that 
always succeeds and leaves the goal unchanged. It is only used to direct the goal to the correct place using the goal types on the output wires. If a goal starts with 
an existential/universal quantifier, a conjunction or implication, then it is sent to the relevant tactic, and the process is repeated. If not, it is sent to the output. Note that an output goal from this strategy is guaranteed not to start with any of the above symbols.

\vspace{10pt}
\noindent \textbf{Limitations.} 
Currently, the GUI navigation is limited in the sense that the user cannot select specific goals to apply or work with
more than one level of graph hierarchy at the same time. Furthermore, nested graph tactics have to be implemented
separately before they can be used, whereas ideally, these could be created in place. More generally, we would like to
be able to configure tactics more easily in the GUI, both tactics provided by the prover and graph tactics. At the
moment only `breadth-first' and `depth-first' search are supported, while a variant of `breadth-first' goal selection is
possible. So, we would like to improve on the evaluation and search strategies and make it easier for users to develop
and plug-in their own strategies. Finally, we would like to improve the debugging facilities to e.g. enable inspection
from a given point in the graph.

\beforesection
\section{Related Work} \label{sec:related}
\aftersection

The graphical part of PSGraph is described using \emph{string diagrams}, whose rewrite theory was formalised in \cite{paper:Dixon:10}
using a particular family of typed digraphs called \textit{open-graphs}. We have elided most details of the underlying formalisation, and refer to \cite{paper:Dixon:10}. We are not claiming to be more expressive compared with tactic languages found in systems such as Isabelle, PVS and Coq. In particular, many syntactic 
goal type properties can be handled by the matching construct of Coq's {\cal L}tac \cite{Delahaye02}. However, we do believe that the way we handle the flow of goals is more natural, and PSGraphs are easier to debug, and may lead to more robust proof strategies, by making users think more about where goals should go next.

Tactics in common theorem provers are essentially untyped (even in {\cal L}tac), meaning there is
limited, if any, support for static checking. However, the idea of ``types'', or goal properties, for tactics, which can be checked 
statically, is not new.  In  \emph{proof planning}~\cite{Bundy91} tactics are given pre-conditions and post-conditions. This entails a significant amount of reasoning just to compose them, thus we have opted for a more light-weight version with our goal types. Moreover, our graphs provide additional flow properties to guide the goals.  
There have also been more type-theoretical approaches to typed tactics,
such as the VeriML language \cite{Stampoulis10}. PSGraph deviates from VeriML by
using (goal) types purely to compose tactics and ensure that goals are sent to the correct target. In VeriML, the types include information about the relationship between tactics and the proofs produced. As the goal of PSGraph is to be theorem prover generic, this is assumed to be property of the theorem prover. In that sense, it is closer to proof planning. In fact, PSGraph did  initially start as a new version of the IsaPlanner proof planner \cite{paper:Dixon:03}, however this was abandoned for pragmatic reasons.
We believe our way of capturing the flow of goals by utilising goal types and 
essentially treating composition as ``piping", is novel for proof (strategy) languages.

When writing proofs, as opposed to proof strategies, one often 
distinguishes between \emph{procedural} proofs, where a proof is described as a sequence of tactic applications (i.e. function composition), ignoring the goals; and \emph{declarative} or \emph{structured} proofs, where the proof is described in terms of intermediate goals (goal islands), and the actual proof  commands are seen more as a side issue.  We can view PSGraph as a marriage of these concepts in the sense that the \emph{goal-type} and goals on the wires create a declarative view, while the graph as a whole gives a \emph{procedural} view of how tactics are composed.
Autexier and Dietrich \cite{Autexier10} have developed \emph{a declarative tactic language on top of a declarative proof language}. Their work is more declarative than PSGraph, whilst our is more general w.r.t. compositions, as they represent a strategy as a \emph{schema} which needs to be instantiated. 
Similarly, there have been several attempts to create \emph{declarative tactic languages on top of procedural tactic languages} \cite{Harrison96,Giero:07}. Asperti et al \cite{Asperti09} argues that these approaches suffer from two
drawbacks: goal selection for multiple sub-goals, and information flow between tactics -- both of these are addressed by goal types in PSGraph. HiTac is a tactic language with additional support for hiding complexities using hierarchies \cite{paper:Aspinall:2008,paper:Whiteside:11}. Graph tactics have been inspired by this work, however the use of goal types on input wires enables multiple
goals as input without introducing non-determinism or relying on goal order, whereas HiTac is restricted to a single input goal.

Finally, it is important to note the difference with the field of \emph{diagrammatic reasoning}, as in e.g. \cite{Jamnik01} and \cite{KissingerThesis}, where diagrams are the objects of interest for reasoning rather than the means of capturing the reasoning process.

\beforesection
\section{Conclusion and Future Work}\label{sec:conc}
\aftersection

We have presented the PSGraph language together together with an implementation of it in the \psgraphtool{} tool. PSGraph's ``lifting" of proof strategies to the level of goal-types, rather than the level of goals,  enables us to write more robust
strategies that no longer rely on the number and order of sub-goals resulting from a tactic application for tactic composition. Moreover, as composition of proof strategies is also at the level of goal-types, we increase type safety and enable better static analysis. Moreover, the problem of goal selection/focus/classification when composing tactics, as highlighted in \cite{Asperti09}, is significantly improved. Graphs naturally represent the flow of goals, and enable graphical inspection of evaluation to improve debugging of proof strategies.

We have already discussed the current tool's limitations. We are currently working on overcoming some of them by
enhancing the GUI and developing new evaluation strategies. One interesting avenue to pursue is to try to implement some
existing larger compound tactics such as `auto' in Isabelle. We suspect that this work will be quite useful in terms
developing goal types that are necessary to direct goals in non-trivial strategies. One way to approach this problem is
to draw the strategies with all goal types being $any$ and use machine learning techniques on a large number of examples
to discover the goal type for each wire. We would also like to develop a notion of sub-typing for goal types, e.g.
anything should be able to be plugged into an $any$ goal type. We are also in the process of starting to use PSGraph to
find new proof strategies by data mining existing libraries as well as for \emph{analogical reasoning}. A first attempt
on using PSGraph for analogical reasoning can be found in \cite{grov13a}. Finally, as we support multiple theorem
provers, it will also be interesting to see if strategies we develop can be carried across theorem provers, thus using
PSGraph as a form of proof (strategy) exchange.

\vspace{-1em}
{\small
\subsubsection*{\footnotesize Acknowledgements} Several of the ideas behind the language is joint with Lucas Dixon, and Alan Bundy provided valuable comments on a
previous version of this paper.  Also thanks to Alex Merry, Rod Burstall, 
Andrius Velykis and Ewen Maclean for suggestions and discussions, and the anonymous   reviewers for constructive feedback. This work has been supported by EPSRC grants: EP/H023852, EP/H024204 and EP/J001058, the John Templeton Foundation, and the Office of Navel Research.
}
\vspace{-1em}

\begin{small}
\bibliographystyle{akbib}
\bibliography{stratlang}
\end{small}

\end{document}